\RequirePackage{fix-cm}
\documentclass{svjour3}                     
\smartqed  

\usepackage{graphicx}
\usepackage{graphicx}
\usepackage{empheq}
\usepackage{subfigure}

\usepackage{tikz-cd}
\usepackage{tikz}

\usepackage[numbers]{natbib}
\usepackage[integrals]{wasysym}
\usepackage{changes}

\usetikzlibrary{decorations.markings}
\usetikzlibrary{shapes.geometric, arrows}
\usepackage{tikzscale}
\usepackage{filecontents}
\usepackage{wrapfig}
\usepackage{tcolorbox}
\usepackage{color}
\definecolor{blue1}{rgb}{0.0, 0.0, 1.0}
\definecolor{gray}{rgb}{0.9,0.9,0.9}
\definecolor{gray1}{rgb}{0.7,0.7,0.7}
\definecolor{gray2}{rgb}{0.8,0.8,0.8}
\definecolor{magenta}{rgb}{1.0, 0.0, 1.0}
\usepackage[title]{appendix}
\newcommand\backmatter{\appendix
\def\chaptermark##1{\markboth{%
\ifnum  \c@secnumdepth > \m@ne  \@chapapp\ \thechapter:  \fi  ##1}{%
\ifnum  \c@secnumdepth > \m@ne  \@chapapp\ \thechapter:  \fi  ##1}}%
\def\sectionmark##1{\relax}}
 
\usepackage{adjustbox}

\newcommand{\ISI}[1]{\mathrm{ISI}}

\usepackage{hyperref}
\hypersetup{ colorlinks=true, urlcolor  = blue, linkcolor = blue,
citecolor = blue1,}

\usepackage{float}
\usepackage{amsmath,mathtools,amssymb,amsfonts}
\usepackage[scaled=.95]{helvet}
\usepackage{multirow}
\usepackage{rotating}

\usepackage[linesnumbered,ruled,vlined]{algorithm2e}
\usepackage[noend]{algpseudocode}
\SetKwInput{KwInput}{Input}                
\SetKwInput{KwOutput}{Output}              

\SetCommentSty{mycommfont}

\newcommand{\marius}[1]{\textcolor{black}{#1}}

\definecolor{lime}{HTML}{A6CE39}
\DeclareRobustCommand{\orcidicon}{%
    \begin{tikzpicture}
    \draw[lime, fill=lime] (0,0) 
    circle [radius=0.16] 
    node[white] {{\fontfamily{qag}\selectfont \tiny ID}};
    \draw[white, fill=white] (-0.0625,0.095) 
    circle [radius=0.007];
    \end{tikzpicture}
    \hspace{-2mm}
}

\foreach \x in {A, ..., Z}{%
    \expandafter\xdef\csname orcid\x\endcsname{\noexpand\href{https://orcid.org/\csname orcidauthor\x\endcsname}{\noexpand\orcidicon}}
}
\newcommand{\orcid}[1]{\href{https://orcid.org/#1}{\textcolor[HTML]{A6CE39}{\orcidicon}}}

\newcommand{\uproman}[1]{\uppercase\expandafter{\romannumeral#1}}

\journalname{X Journal}

\begin{document}

\title{Synchronization in STDP-driven memristive neural networks with time-varying topology}

\titlerunning{Synchronization in memristive neural networks with time-varying topology}        

\author{Marius E. Yamakou\orcid{0000-0002-2809-1739}, Mathieu Desroches\orcid{0000-0002-9325-4207}, and Serafim Rodrigues\orcid{0000-0002-3601-5760}}


\institute{M. E. Yamakou \at Department of Data Science, Friedrich-Alexander-Universit\"at Erlangen-Nürnberg, Cauerstr. 11, 91058 Erlangen, Germany\\
M. E. Yamakou \at Max-Planck-Institut f\"ur Mathematik in den Naturwissenschaften, Inselstr. 22, 04103 Leipzig, Germany\\
M. Desroches \at MathNeuro Project-Team, Inria Center at Universit{\'e} C{\^o}te d'Azur, 2004 route des Lucioles - BP 93, 06902 Sophia Antipolis Cedex, France\\
Serafim Rodrigues  \at Mathematical, Computational and Experimental Neuroscience, Basque Center for Applied Mathematics, Alameda de Mazzaredo 14, 48009 Bilbao, Spain\\
Serafim Rodrigues  \at Ikerbasque, Basque Foundation for Science, Plaza Euskadi 5, 48009 Bilbao, Spain\\
             \email{marius.yamakou@fau.de}\\  
}

\date{Received: date / Accepted: date}

\maketitle
\begin{abstract}
  Synchronization is a widespread phenomenon in the brain. Despite numerous studies, the specific parameter configurations of the synaptic network structure and learning rules needed to achieve robust and enduring synchronization in neurons driven by spike-timing-dependent plasticity (STDP) and temporal networks subject to homeostatic structural plasticity (HSP) rules remain unclear. Here, we bridge this gap by determining the configurations required to achieve high and stable degrees of complete synchronization (CS) and phase synchronization (PS) in time-varying small-world and random neural networks driven by STDP and HSP.  In particular, we found that decreasing $P$ (which enhances the strengthening effect of STDP on the average synaptic weight) and increasing $F$ (which speeds up the swapping rate of synapses between neurons) always lead to higher and more stable degrees of CS and PS in small-world and random networks, provided that the network parameters such as the synaptic time delay $\tau_c$, the average degree  $\langle k \rangle$, and the rewiring probability $\beta$ have some appropriate values. When $\tau_c$, $\langle k \rangle$, and $\beta$ are not fixed at these appropriate values, the degree and stability of CS and PS may increase or decrease when $F$ increases, depending on the network topology. It is also found that the time delay $\tau_c$ can induce intermittent CS and PS whose occurrence is independent $F$. Our results could have applications in designing neuromorphic circuits for optimal information processing and transmission via synchronization phenomena.
\end{abstract}

\section{Introduction}\label{Sec. I}
Synchronization phenomena are processes wherein many dynamical systems adjust a given property (e.g., amplitude, phase, frequency, and even membrane potential in coupled neurons) of their motion due to suitable coupling configurations. In the brain, they can emerge from the collaboration between neurons or neural networks and significantly affect all neurons and network functioning. It is well-established that synchronization of neural activity within and across brain regions promotes normal physiological functioning, such as the precise temporal coordination of processes underlying cognition, memory, and perception \cite{neustadter2016eeg}. However, synchronization of neural activity is also well known to be responsible for some pathological behaviors such as epilepsy \cite{lehnertz2009synchronization}. \marius{It has been shown that changes in the strength of the synaptic coupling and the connectivity of the neurons could lead to epileptic-like synchronization behaviors. Furthermore, changes in neural connectivity can lead to hyper-synchronized states related to epileptic seizures that occur intermittently with asynchronous states \cite{borges2023intermittency}.  It has been demonstrated in \cite{protachevicz2019bistable} that by manipulating synaptic coupling and creating a hysteresis loop, square current pulses can induce abnormal synchronization similar to epileptic seizures.} Synchronization may present various forms (see \cite{boccaletti2006complex,osipov2007synchronization} for a comprehensive review), and the behavior of each form of synchronization may depend on the nature of the interacting systems, the type of coupling, the distances between the interacting systems, the time delays between the components of the systems, and also the network topology. 

In this paper, we focus on two common forms of synchronization for reasons given alongside their descriptions: (i) Complete synchronization (CS) is the simplest (and probably the most intuitive) form of synchronization. A system made up of, e.g., two coupled sub-systems, say $x_1(t)$ and $x_2(t)$, is said to be completely synchronized when there is a set of initial conditions so that the coupled systems eventually evolve identically in time (i.e., $\lvert x_1(t) - x_2(t)\rvert=0$, as $t \to \infty$) \cite{osipov2007synchronization,fujisaka1983stability,pecora2015synchronization,yamakou2016ratcheting}. Because of the intuitiveness and simplicity of CS, it will be one of the main phenomena investigated in this paper.  (ii) Phase synchronization (PS) was introduced by Rosenblum et al. \cite{rosenblum1996phase,pikovsky1996synchronization} and experimentally confirmed in \cite{parlitz1996experimental}. It involves sub-system properties called phases \cite{pietras2019network} and is characterized by $2\pi$ phase locking of two or more oscillators with uncorrelated amplitudes. It has been shown that the phase synchronization between different brain regions supports both working memory and long-term memory and facilitates neural communication by promoting neural plasticity \cite{fell2011role}, making PS a good candidate for investigation in this paper.

In recent years, extensive research (see, e.g., the reviews in \cite{tang2014synchronization,boccaletti2002synchronization,arenas2008synchronization}) has been conducted on synchronization dynamics in non-adaptive neural systems with varying degrees of complexity. \marius{In particular, in the study \cite{protachevicz2021emergence}, it was discovered that excitatory and inhibitory connections between brain areas are crucial for phase and anti-phase synchronization. It was found that the phase angles of neurons in the receiving area could be influenced by unidirectional non-adaptive synapses from the sender area. When the neurons in the sender area synchronize, the variability of phase angles in the receiver area can be reduced with certain conductance values. Additionally, the study observed both phase and anti-phase synchronization in the case of non-adaptive bidirectional interactions. It has also been demonstrated in \cite{borges2017synchronised} that the coupling strength and the probability of connections in a random network of adaptive exponential integrate-and-fire neurons can induce spike and bursting synchronization, with bursting synchronization being more robust than spike synchronization.}

\marius{Furthermore, it has been shown that axonal time delays can play crucial roles in synchronization dynamics in neural networks \cite{hansen2022effect,khoshkhou2019spike,lameu2018alterations,madadi2023delay,protachevicz2020influence}. 
For example, in \cite{madadi2023delay}, the authors used phase oscillator and conductance-based neuron models to study synchronization and coupling between two bidirectionally coupled neurons in the presence of transmission delays and STDP, which influence emergent pairwise activity-connectivity patterns. Their results showed that depending on the range of transmission delays, the two-neuron motif could achieve in-phase/anti-phase synchronization and symmetric/asymmetric coupling. The co-evolutionary dynamics of the neuronal system and synaptic weights, governed by STDP, stabilize the motif in these states through transitions at specific transmission delays. They further showed that these transitions are sensitive to the phase response curve of the neurons but are robust to heterogeneity in transmission delays and STDP imbalance. Motivated by such rich time-delay-induced dynamical behavior in synchronization dynamics, in the current paper, we shall investigate the effects of axonal time delays on CS and PS in neural networks driven by two forms of adaptive rules.}

It is essential to consider the effects of the inherently adaptive nature of neural networks on information processing via synchronization. \marius{Besides the colossal efforts to study synchronization in neuronal networks with synaptic plasticity, see, e.g., \cite{lameu2018alterations,protachevicz2023plastic,ratas2022interplay,schmalz2019controlling,silveira2021effects,solis2021model}, it is essential to be mindful of the need to explore more dynamic scenarios in order to fully comprehend the emergence of synchronous patterns in adaptive networks.} \marius{Synaptic plasticity} in neural networks refers to the ability to modify the strength of synaptic couplings over time and/or the architecture of neural network topology through specific rules. Two significant mechanisms associated with adaptive rules in neural networks are spike-timing-dependent plasticity (STDP) and homeostatic structural plasticity (HSP). \marius{STDP-induced synaptic modification relies on the repeated pairing of pre-and postsynaptic membrane potentials. The degree and direction of the modification depend on the relative timing of neuron firing.}  Depending on the precise timing of pre-and and postsynaptic spikes, the synaptic weights can either exhibit long-term depression (LTD) or long-term potentiation (LTP), which represent persistent weakening or strengthening of synapses, respectively. This concept has been extensively discussed in \cite{gerstner1996neuronal,markram1997regulation}. 

\marius{HSP-induced synaptic modification involves altering the connectivity between neurons by creating, pruning, or swapping synaptic connections. This results in changes to the network's architecture while maintaining its functional structure, which maximizes specific functions of interconnected groups of neurons and improves sensory processing efficiency \cite{shine2016dynamics}. Early evidence of structural plasticity was observed through histological studies of spine density following new sensory experiences or training~\cite{greenough1988anatomy}. Further research has shown that the micro-connectome, which describes the connectome at the level of individual synapses, undergoes rewiring \cite{bennett2018rewiring,van2017rewiring,yamakou2023combined}. While brain networks adhere to specific topologies, such as small-world and random networks \cite{hilgetag2016brain,valencia2008dynamic}, despite their time-varying dynamics, recent studies suggest that these networks can benefit from homeostasis by increasing the efficiency of information processing \cite{butz2014homeostatic}. Motivated by these studies, the current paper focuses on time-varying small-world and random networks adhering to their respective topologies through HSP.}

Previous studies \cite{borges2016effects,borges2017synaptic,kim2018effect,kim2018effect1} on synchronization in adaptive neural networks have focused on either time-invariant neural networks with STDP or time-varying neural networks without STDP. Research on time-invariant neural networks has shown that good synchronization improves via LTD of the averaged synaptic weight, while bad synchronization deteriorates via LTP \cite{kim2018effect}. This effect is due to inhibitory STDP \cite{kim2018effect}, which contrasts the findings on excitatory STDP \cite{kim2018stochastic}, where good synchronization gets better and bad synchronization gets worse via LTP and LTD, respectively.  The article \cite{talathi2008spike} demonstrated that STDP enhances synchronization in inhibitory networks even when there is heterogeneity. Similarly, \cite{popovych2013self} revealed that noise can facilitate synchronization in spiking neural networks driven by STDP. It is shown that the average synaptic coupling of the network increases with an increase in the noise intensity, with an optimal noise level where the strength of average synaptic coupling reaches its maximum in a resonance-like fashion that maximizes synchronization. 
The research in \cite{borges2016effects} demonstrated the crucial combined effect of the uni-directional chemical synapses and STDP on the synchronization in random neural networks. The study also reveals that synchronization increases as the connection probability of the network grow in the presence of STDP and no external input current.

However, introducing a non-zero external input current results in spiking resynchronization. The study in \cite{ren2007adaptive} explores the behavior of an adaptive array of phase oscillators and highlights that a specially designed adaptive law can amplify the coupling between pairs of oscillators with greater phase incoherence, leading to improved synchronization. This approach yields more realistic coupling dynamics in networks of oscillators with varying intrinsic frequencies. Additionally, adjusting the parameters of the adaptive law can accelerate synchronization. The paper also demonstrated the method's versatility by examining nearest-neighbor ring coupling in addition to global coupling.

The research in \cite{faggian2019synchronization,ghosh2022synchronized,rakshit2018emergence} has shown that in networks with a time-varying topology but without STDP, a faster rewiring of the topology always leads to a higher degree of synchronization. However, in our current work, we challenge this notion and demonstrate that more rapid switching of synapses can actually also decrease the degree of synchronization in certain situations. The issues of synchronization phenomena in networks undergoing two adaptive processes have not received sufficient research attention. In one study, published in \cite{chauhan2022dynamics}, the authors examined this problem by analyzing Kuramoto oscillator networks that undergo two adaptation processes: one that modifies coupling strengths and another that changes the network structure by pruning existing synaptic contacts and adding new ones. By comparing networks with only STDP to those with both STDP and structural plasticity, the authors assessed the effects of structural plasticity and found that it enhances the synchronized state of a network.

The current study aims to narrow the gap in the research on synchronization in time-varying neural networks driven by STDP and HSP rules in small-world and random networks. Specifically, we focus on determining: (i) the joint effect of the adjusting potentiation rate of the STDP rule and the characteristic rewiring frequency of the HSP rule on the degree of CS and PS, (ii) the joint effect of the synaptic time delay, the rewiring frequency of the HSP rule, and the adjusting potentiation rate of the STDP rule on the degree of CS and PS, (iii) the joint effect of the average degree of the network, the rewiring frequency of the HSP rule, and the adjusting potentiation rate of the STDP rule on the degree of CS and PS, and (iv) the joint effect of the rewiring probability of the Watts-Strogatz small-world network, the rewiring frequency of the HSP rule, and the adjusting potentiation rate of the STDP rule on the degree of CS and PS. The study employs extensive numerical simulations to investigate these issues.

Based on our numerical results, the stability of degrees of CS and PS are influenced by parameters governing STDP and HSP, as well as network topology parameters. For instance, decreasing the STDP potentiation rate parameter ($P$) and increasing the HSP characteristic frequency parameter ($F$) leads to more stable and higher levels of CS and PS in small-world and random networks, provided that average degree ($\langle k \rangle$), rewiring probability ($\beta$), and synaptic time delay ($\tau_c$) are at appropriate values. Furthermore, we found that PS can be achieved more reliably and at a higher degree than CS in both small-world and random networks. Additionally, the random network generates more stable and higher levels of CS and PS than the small-world network. Our findings on the variations in the degree of CS and PS are summarized in Table \ref{tab1}.

The paper is structured as follows: Sec. \ref{Sec. II} describes the mathematical model, the STDP learning rule, and the HSP rewiring rules, which facilitate the adherence of time-varying small-world and random networks to their respective architecture. Sec. \ref{Sec. III} outlines the computational methods utilized, while \ref{Sec. IV} presents and analyzes the numerical findings. In Sec. \ref{Sec. V}, we have conclusions. 

\section{Model description}\label{Sec. II}
\subsection{Neural Network Model}
The presence of intracellular and extracellular ions leads to the development of an electromagnetic field in biological neurons, which affects their membrane potential and, consequently, their firing modes. To incorporate these effects in a memristive neuron model, Lv \textit{et al.} \cite{lv2016model} proposed improved neuron models that include a variable for magnetic flux. The influence of this electromagnetic field is well-established \cite{lv2016multiple}. Thus, in the current work, we study the joint effects of HSP and STDP on synchronization in a memristive neural network.
\marius{The FitzHugh-Nagumo (FHN) model \cite{fitzhugh1961impulses,nagumo1962}, initially proposed to describe the spiking activity of neurons, now serves as a fundamental model for excitable systems. Its applications have expanded beyond neuroscience and biological processes \cite{ciszak2003anticipating} to include optoelectronics \cite{rosin2011pulse}, chemical oscillators \cite{shima2004rotating}, and nonlinear electronic circuits \cite{heinrich2010symmetry}. Although the FHN model lacks the same level of biophysical relevance as the Hodgkin-Huxley (HH) neuron model \cite{hodgkin1952quantitative}, it nevertheless does capture some essential aspects of the HH model's behavior. Moreover, the computational cost is reduced due to the lower dimensionality of the 2D FHN model compared to the 4D HH model, which is particularly advantageous when analyzing large networks. Our study considers the memristive FHN model, incorporating the memristive aspect via an additional equation as per \cite{fu2018subcritical}:}
\begin{eqnarray}\label{eq:1}
\left\{\begin{array}{lcl}
\displaystyle{\frac{dv_i}{dt}}&=&v_i\big(v_i-a\big)\big(1-v_i\big)-w_i + k_3v_i\rho(\phi_i)- I_{i}^{syn}(t),\\[2.0mm]
\displaystyle{\frac{dw_i}{dt}}&=&\varepsilon\big(v_i-dw_i\big),\\[3.0mm]
\displaystyle{\frac{d\phi_i}{dt}}&=&k_1v_i - k_2\phi_i + \phi_{ext},
\end{array}\right.
\end{eqnarray}
where the variables $v_i$, $w_i$, and $\phi_i$ correspond to the voltage, slow current variable, and magnetic flux, respectively. To maintain electrophysiological relevance, the parameter $a$ is typically set within the $(0,1)$ range, with a $0.5$ chosen for our purposes \cite{xu2014parameters}. The values of $\varepsilon$ and $d$ are fixed at $0.025$ and $1$, respectively, representing a specific set of values at which the non-memristive FHN model (i.e., Eq.\eqref{eq:1} with $k_1=K_2=k_3=0$) is at the quiescent state. The flux-controlled memristor term $\rho(\phi_i)$ is modeled using $\rho(\phi) = \lambda+3\beta\phi^2$, where $\lambda$ and $\beta$ are fixed at $0.1$ and $0.02$, respectively \cite{yamakou2020chaotic}. The memristor parameters $k_1=0.5$, $k_2=0.9$, $k_3=1.0$, and $\phi_{ext}=2.4$ are also fixed. \marius{With these parameter  values, the model in Eq.\eqref{eq:1} can only produce regular spiking \cite{fu2018subcritical} --- the regime in which we are interested.}

\subsection{Synapses and STDP Rule}
The term $I_{i}^{syn}(t)$ in Eq.\eqref{eq:1} represents the uni-directional excitatory chemical synapses between neurons and governs the STDP learning rule between coupled neurons. The synaptic current $I_{i}^{syn}(t)$ of the $i$th neuron at time $t$ is defined in Eq.\eqref{eq:5}:
\begin{equation}\label{eq:5}
I_{i}^{syn}(t) = \frac{1}{k_i}\sum_{j=1(\neq i)}^{N}\ell_{ij}(t)g_{ij}(t)s_j(t)\big[v_i(t)- v_{syn}\big],
\end{equation}
where the synaptic connectivity matrix $L (=\{\ell_{ij}(t)\})$ has $\ell_{ij}(t)=1$ if neuron $j$ is connected to neuron $i$ and disconnected when $\ell_{ij}(t)=0$. We model the synaptic connections as either a time-varying small-world network or a time-varying random network. Starting with a regular ring network with $\langle k \rangle $ nearest neighbors, we use the Watts-Strogatz algorithm \cite{watts1998collective} to generate small-world and random networks with parameters $\beta$ and $\langle k \rangle$, where $\beta$ represents the rewiring probability and ranges from 0 to 1, and $\langle k \rangle$, the average degree connectivity (i.e., the average number of synaptic inputs per neuron), which is calculated as $\langle k \rangle =\frac{1}{N} \sum_{i=1}^{N}k_i$, where $k_i$ is the in-degree of the $i$th neuron (i.e., the number of synaptic inputs to neuron $i$) and is given by $k_i=\sum_{j=1(\neq i)}^{N}\ell_{ij}(t)$. In the algorithm, $\beta\in[0,1]$ plays a crucial role in determining the type of network generated. If $\beta$ falls between 0 and 1, a small-world network is created, while a completely random network is generated when $\beta$ is 1.  This work does not consider regular networks (when $\beta$ is 0). The average degree connectivity $\langle k \rangle$ and the rewiring probability $\beta$ serve as control parameters for the network topology.

The time-dependent behavior of the open synaptic ion channels in the $j$th neuron is denoted by $s_j(t)$ in Eq.\eqref{eq:5}. The rate of change of $s_j(t)$ is determined by
 \begin{equation}\label{eq:5a}
\frac{ds_j}{dt} = \frac{2(1 - s_j)}{1 + \displaystyle{\exp\bigg[- \frac{v_j(t-\tau_c)}{v_{shp}}\bigg]}}-s_j.
 \end{equation}
\marius{Chemical synapses involve the release and diffusion of neurotransmitters across the synaptic cleft, which takes a finite amount of time. Including time delays allows for a more accurate representation of the temporal dynamics and signal transmission between neurons. Thus, we incorporate a time delay parameter, $\tau_c$, which will be utilized to control the chemical synapses. With a time delay $\tau_c$,  the action potential of the pre-synaptic neuron $j$ fired at the earlier time given by $t-\tau_c$ is represented by $v_j(t-\tau_c)$ \cite{madadi2023delay,yu2015spike}.}  The threshold of the membrane potential, denoted by $v_{shp}=0.05$, determines the threshold above which the pre-synaptic neuron $j$ has an impact on post-synaptic neuron $i$. Additionally, the reversal potential, set at $v_{syn} = 2.0$, ensures that all synapses are excitatory.
 
In Eq.\eqref{eq:5}, the strength of the synaptic connection between the $j$th pre-synaptic neuron and the $i$th post-synaptic neuron is denoted by $g_{ij}(t)$. The STDP mechanism states that the synaptic strength of each synapse is updated using a nearest-spike pair-based STDP rule \cite{morrison2007spike} as time $t$ increases.
\marius{There are two commonly used forms of STDP, see, e.g., \cite{lameu2021short,protachevicz2023plastic,yang2022chimera} and \cite{song2000competitive,xie2018spike,yu2014effects}, for each of the forms.} 
In our study, the update of the synaptic coupling strength $g_{ij}(t)$ is determined by the synaptic modification function $M$, which is defined based on the current value of $g_{ij}(t)$ \cite{song2000competitive,xie2018spike,yu2014effects}:
\begin{eqnarray}\label{eq:6}
\left\{\begin{array}{lcl}
 g_{ij}(t + \Delta t) = g_{ij}(t) + \Delta g_{ij},\\[3.0mm]
\Delta g_{ij}=g_{ij}(t)M(\Delta t),
  \\[3.0mm]
M(\Delta t)=
  \left\{
\begin{array}{ll}
\displaystyle{P\exp{(-\lvert\Delta t\rvert/\tau_{p})}\:\:\text{if}~\Delta t>0}\\[1.0mm]
\displaystyle{- D\exp{(-\lvert\Delta t\rvert/\tau_{d})}\:\:\text{if}~\Delta t<0}\\[1.0mm]
\displaystyle{0 \:\:\text{if}~\Delta t=0,}
\end{array} 
\right.
\end{array}\right.
\end{eqnarray}
where $\Delta t=t_i -t_j$, with $t_i$ and $t_j$ representing the spiking times of the post-synaptic neuron $i$ and the pre-synaptic neuron $j$, respectively. We determine the spike occurrence times from the instant $t$ when a membrane potential variable crosses the threshold value of $v_{\mathrm{th}}=0.5$.
\marius{It is worth noting that only the excitatory-to-excitatory synapses are modified by this learning rule \cite{li2009self,song2000competitive}, making it an ideal learning rule for our study since all the synapses in our network are excitatory --- thanks to the value of the reversal potential, $v_{syn} = 2.0$, which ensures that all synapses are excitatory. The extent of synaptic modification is regulated by two parameters, namely the potentiation and depression rate represented by $P$ and $D$, respectively. The temporal window for synaptic modification is determined by two additional parameters, $\tau_p$ and $\tau_d$. Experimental results \cite{bi1998synaptic,feldman2005map,song2000competitive} suggest that $D\tau_d>P\tau_p$, which ensures the overall weakening of synapses. Furthermore, experimental studies show that the temporal window for synaptic weakening is roughly the same as that for synaptic strengthening \cite{song2000competitive,zhang1998critical}. Hence, to be consistent with experimental results, we chose the STDP parameters such that the STDP rule in Eq.\eqref{eq:6} is typically depression-dominated, i.e., we set $\tau_p=\tau_d=2.0$, $D/P=1.05$, and chose $P$ as the control parameter of this STDP rule.} In order to prevent unbounded growth, negative coupling strength, and elimination of synapses (i.e., $g_{ij}=0$), we set a range with the lower and upper bounds: $g_{ij}\in[g_{min},g_{max}]=[0.001,0.5]$.
\subsection{Time-varying Networks and HSP Rule}
\marius{To investigate the impact of the time-varying nature of the network architectures on the synchronization dynamics of the coupled neurons, we consider a small-world and random structure \cite{bassett2006small,bassett2006adaptive,liao2017small,muldoon2016small} constructed using a Watts-Strogatz network algorithm \cite{watts1998collective}. The network's Laplacian matrix is a zero-row-sum matrix with an average degree connectivity of $\langle k\rangle$ and a rewiring probability $\beta\in(0,1]$. To generate a time-varying small-world network (with $\beta\in(0,1)$) that adheres to its small-worldness at all times, we implement the following process during the rewiring of synapses:}
\begin{itemize}
\item \marius{\textit{During each integration time step $dt$, a synapse between two distant neurons is rewired to a nearest neighbor of one of the neurons with probability $(1 - \beta)Fdt$. If the synapse is between two nearest neighbors, it is replaced by a synapse to a randomly chosen distant neuron with probability $\beta Fdt$.  A neuron $i$ is considered a distant node to neuron $j$ if $\lvert i-j \rvert >\langle k \rangle$, where $\langle k \rangle$ is the average degree of the original ring network used in the Watts-Strogatz algorithm.}}
\end{itemize}
\marius{To generate a time-varying random network (also generated with the Watts-Strogatz algorithm when $\beta=1$) that adheres to its randomness at all times, we implement the following process during the rewiring of synapses:}
\begin{itemize}
\item \marius{\textit{During each integration time step $dt$, if there is a synapse between neuron $i$ and $j$, it will be rewired such that neuron $i$ ($j$) connects to any other neuron except for \marius {neuron} $j$ ($i$) with a probability of $\big(1-\frac{\langle k\rangle}{N-1}\big)Fdt$.}}
\end{itemize}

\marius{Note that the rewiring algorithms described above always maintain the small-worldness or randomness of the networks, even though the connectivity matrix changes over time --- these are precisely the HSP rules we will use in this study. However, it is essential also to acknowledge that real neural networks may employ different rewiring processes to achieve such time-varying network structures, which may not necessarily align with the HSP rules described here. Nonetheless, for the purpose of our study, it is relevant that both small-world and random networks exhibit changing connections over time while preserving their respective small-worldness or randomness, similar to what is observed in real neural networks. 
}

Here, we will use the characteristic rewiring frequency $F$ as the control parameter for HSP. This parameter reflects the synapse changes over time, specifically during each integration time step $dt$. Notably, synapses in actual neural networks may change at varying rates, depending on factors such as the network's developmental stage or environmental stimuli. Therefore, this study aims to investigate a broad range of rewiring frequencies, ranging from 0.0 to $1.0\times10^2$.

\section{Computational Methods}\label{Sec. III}
As we need to quantify the degree of complete synchronization (CS) and phase synchronization (PS) of neural activity in the networks, we use the error for variable traces $E$ for CS and the Kuramoto order parameter $R$ for PS \cite{bertolotti2017synchronization,kuramoto1984chemical}, respectively given by
\begin{equation}\label{eq:7}
\left\{\begin{array}{lcl}
\displaystyle{E=\Bigg \langle\frac{1}{N-1}\sum\limits_{i=2}^N\sqrt{(v_i-v_1)^2+(w_i-w_1)^2+(\phi_i-\phi_1)^2}\Bigg \rangle_{{t}}},\\[4.0mm]
\displaystyle{R=\Bigg \langle\bigg\lvert \frac{1}{N}\sum\limits_{i=1}^N\exp{\big[z\Psi_{i}(t)\big]}\bigg\rvert\Bigg \rangle_{{t}}},
\end{array}\right.
\end{equation}
where $\Psi_{_{i}}(t)= 2\pi \ell +2\pi \frac{ t-t_i^{(\ell)}}{t_i^{(\ell+1)}-t_i^{(\ell)}}$, $ t_i^{(\ell)}\leq t<t_i^{(\ell+1)}$, and $\big \langle \cdot \big\rangle_{{t}}$ is the time average obtained over a large time interval $[0,T]$.  
In the argument of the exponential function, we have $z = \sqrt{-1}$, and the quantity $\Psi_{_{k}}(t)$ approximates the phase of the $k$th neuron and linearly increases over $2\pi$ from one spike to the next.
We determine the spike time occurrences from the instant the membrane potential variable $v_k$ crosses the threshold $v_{\mathrm{th}}=0.5$ from below. The norm of this complex exponential function is represented by  $\lvert \cdot \rvert$. The time at which the $i$th neuron fires its $\ell$th spike ($\ell = 0, 1, 2, ...$) is represented by $t_i^{(\ell)}$. 

CS corresponds to when all neurons follow the same trajectory and yields zero synchronization error $E=0$. The Kuramoto order parameter \textbf{$R$} ranges from 0 to 1, corresponding to the \marius{absence of PS to complete PS} (i.e., all neurons fire at precisely the same times), respectively. It is worth noting that the error $E$, which measures the degree of CS, uses the actual and all the values of the membrane variable $v_{_{k}}(t)$ (including subthreshold oscillations), while the Kuramoto order parameter uses only the spike times of $v_{_{k}}(t)$ to inform us about the synchronization of spiking times. Thus, the synchronization behavior of the neurons during CS can be very different from what happens during PS. 

For $N$ neurons, we numerically integrate Eqs.\eqref{eq:1}-\eqref{eq:5a} with the STDP learning rule of Eq.\eqref{eq:6} and the HSP \marius{rewiring models} described above using a standard fourth-order Runge-Kutta algorithm with a time step $dt = 0.01$ and for a \marius{total integration time of $T=3.0\times10^{3}$ units.} The results are shown in Section~\ref{Sec. IV} below were averaged over 25 independent realizations for each set of parameter values and random initial conditions to warrant reliable statistical accuracy with respect to the small-world and random network generations and the global stability of CS and PS.
For each realization,  we choose random initial points $[v_k(0),w_k(0),\phi_k(0)]$ for the $k$th ($k = 1,...,N$) neuron with uniform probability distribution in the range of $v_k(0)\in[-0.5, 1.6]$, $w_k(0)\in[0.1,1]$, $\phi_{k}(0)=[2.45,3.5]$.  It is worth pointing out that we have carefully excluded the transient behavior from simulations as with all the quantities calculated. \marius{After an initial transient time of $T_0=2.4\times10^{3}$ units, we start recording the values of the variables $(v_k,w_k,\phi_k)$ and the spiking times $t_k^{\ell}$ ($\ell\in\mathbb{N}$ counts the spiking times).} Furthermore, the initial weights of all excitable synapses are normally distributed in the interval $[g_{min},g_{max}]=[0.001,0.5]$, with mean $g_0=0.35$ and standard deviation $\sigma_0=0.01$.

The flow of control in the simulations is presented in Table \ref{tab2} and the Algorithm in Appendix. The two outermost loops in the pseudo-code are
on the parameters $P$ and $F$, resulting in Fig. \ref{fig:1}. Other parameters replace the parameter in the outermost loop (i.e., $P$) to get results presented in
the rest of the figures.

The global stability of CS and PS is analyzed using basin stability measure  $B$, defined as
\begin{equation}\label{eq:8}
B=\int_{\Omega}h(\omega)g(\omega)d\omega,
\end{equation}
where $\Omega$ represents the set of all possible random perturbations $\omega$ and $h(\omega)$ equals unity if the neural network converges to synchronized states after a perturbation $\omega$ and zero otherwise.
The density of the perturbed states, represented by $g(\omega)$, satisfies the condition $\int_{\Omega}g(\omega)d\omega=1$. 

In our computation, we integrate the system for a sufficiently large number $Q$ of realizations. Each realization is executed with random initial conditions drawn uniformly from a prescribed region of phase
space. If $q$ is the number of initial conditions that eventually arrive at the synchronous state, then the basin stability $B$ for the synchronous
state is estimated as $q/Q$.  Thus, $B$ is bounded in the unit interval [0,1], whereby $B = 0$ indicates that the synchronized state is completely unstable and has the size of its basin of attraction tending to zero; and when $B = 1$, all sampled initial conditions are pulled to the synchronized state, implying a globally stable synchronized state; and when $0<B<1$, the probability (in the classical sense) of getting the synchronous states for random initial conditions located in the prescribed region of the phase space. We can also interpret $0<B<1$ as the coexistence of synchronized and desynchronized states within a given region of phase space. 

As we indicated earlier, a full level synchronization is hardly attained in many real-world systems \cite{rosenblum1997phase}, including biological neurons, where we can have heterogeneous initial conditions and coupling strengths (which are controlled by STDP) and/or the presence of uncorrelated random perturbations. Even though their degree of synchronization could be very high (i.e., $E\leq\delta$, $0<\delta\ll1$ or $R\leq\delta_0$, $0\ll\delta_0<1$, as $t \to \infty$), it is hardly full (i.e., it is hard to get exactly $E=0$ and $R=1$, as $t \to \infty$). Thus, in our computations, we sample the phase space volume prescribed above and consider $E < 10^{-1}$ and $R>0.9$ a satisfactory precision for CS and PS, respectively.
In the rest of this paper, we use the notations $B^{E}$ and $B^{R}$ to distinguish between the basin stability measure of CS and PS, respectively.

\section{Results}\label{Sec. IV}
The purpose of our study is to examine the impact of the HSP, which is governed by the rewiring frequency parameter $F$, in conjunction with (i) the STDP, which is influenced by the adjusting rate parameter $P$, (ii) the time delay $\tau_c$, (iii) the average degree connectivity $\langle k \rangle$, and (iv) the rewiring probability $\beta$, on the degree of CS and PS in small-world and random networks. Our findings on the alterations of the degree of CS and PS are summarized in Table \ref{tab1}.
\begin{table*}
\begin{minipage}{\textwidth}
\caption{Summary of the relevant combined effects of $P$, $F$, and network parameters on the degree of CS and PS. 
The inclined arrow $\nearrow$ or $\searrow$ represents an increase or a decrease, respectively, in the parameter value in the interval indicated and the degree of synchronization.  The vertical arrow $\uparrow$ or $\downarrow$ indicates, respectively, that the high  or low degree of synchronization stays high or low as the parameters are varied.}\label{tab1}
\begin{center}
\begin{tabular}{|c|c|c|c|c|c|}
\hline
\textbf{Topology}   & \textbf{STDP parameter} & \textbf{Network parameters}& \textbf{HSP parameter} & \textbf{Degree of CS} & \textbf{Degree of PS}  \\
\hline
\multirow{12}{*}{\begin{turn}{90}\textbf{Small-world}\end{turn}}   
& $P\:\searrow$ $(10^{-6},10^{-3}]$ &  $\tau_c=3$, $\langle k \rangle=10$, $\beta=0.25$ & $F\in[0,100]$ &  $ E\:\nearrow$ &  $ R\:\nearrow$ \\ \cline{2-6}
& $P\:\nearrow$ $(10^{-6},10^{-3}]$ &  $\tau_c=3$, $\langle k \rangle=10$, $\beta=0.25$ & $F\in[0,100]$ &  $ E\:\searrow$ &  $ R\:\searrow$ \\ \cline{2-6}
& $P=10^{-6}$  & $\tau_c=3$, $\langle k \rangle=10$, $\beta=0.25$ & $F\:\searrow$ $[0,100]$    &  $E\:\searrow$ & $R$ $\uparrow$ \\ \cline{2-6}
& $P=10^{-6}$  & $\langle k \rangle=10$, $\beta=0.25$, $\tau_c\:\searrow$ $[0,20)$ &  $F\in[0,100]$  &  $ E\:\nearrow$ &  $ R\:\nearrow$\\ \cline{2-6}
& $P=10^{-6}$  & $\langle k \rangle=10$, $\beta=0.25$, $\tau_c\in[20,60)$ &  $F\in[0,100]$  &  $ E\:\downarrow$ &  $ R\:{\downarrow}$\\ \cline{2-6}
& $P=10^{-6}$  & $\langle k \rangle=10$, $\beta=0.25$, $\tau_c\:\searrow$ $[60,80]$ &  $F\in[0,100]$  &  $ E\:\nearrow$ &  $ R\:\nearrow$\\ \cline{2-6}
& $P=10^{-6}$   &  $\tau_c=3$,  $\beta=0.25$, $\langle k \rangle\:\nearrow$  $[2,20]$  &  $F\in[0,100]$   &  $E\:\nearrow$ & $R\:\nearrow$  \\ \cline{2-6}
& $P=10^{-6}$   &  $\tau_c=3$,  $\beta=0.25$, $\langle k \rangle\:\searrow$  $[2,20]$  &  $F\in[0,100]$   &  $E\:\searrow$ & $R\:\searrow$  \\ \cline{2-6}
& $P=10^{-6}$   &  $\tau_c=3$,  $\beta=0.25$, $\langle k \rangle\in[2,5]$  &  $F\:\searrow$ $[0,100]$   &  $ E\:\searrow$ &  $R\:\searrow$ \\ \cline{2-6}
& $P=10^{-6}$   &  $\tau_c=3$,  $\beta=0.25$, $\langle k \rangle\in[2,5]$  &  $F\:\nearrow$ $[0,100]$   &   $E\:\nearrow$ & $  R\:\nearrow$ \\ \cline{2-6}
& $P=10^{-6}$   &  $\tau_c=3$, $\langle k \rangle=5$, $\beta\:\nearrow$ $[0.05,1)$  &  $F\in[0,1]$   &   $E\:\nearrow$ & $  R\:\nearrow$ \\ \cline{2-6}
& $P=10^{-6}$   &  $\tau_c=3$, $\langle k \rangle=5$, $\beta\:\nearrow$ $[0.05,1)$  &  $F\in[1,100]$   &   $E\:\searrow$ &  $R\:\searrow$  \\ \cline{2-6}
                                
\hline
\multirow{12}{*}{\begin{turn}{90}\textbf{Random}\end{turn}}   
& $P\:\searrow$ $(10^{-6},10^{-3}]$ &  $\tau_c=3$, $\langle k \rangle=10$, $\beta=0.25$ & $F\in[0,100]$ &  $ E\:\nearrow$ &  $ R\:\nearrow$ \\ \cline{2-6}
& $P\:\nearrow$ $(10^{-6},10^{-3}]$ &  $\tau_c=3$, $\langle k \rangle=10$, $\beta=0.25$ & $F\in[0,100]$ &  $ E\:\searrow$ &  $ R\:\searrow$ \\ \cline{2-6}
& $P=10^{-6}$  & $\tau_c=3$, $\langle k \rangle=10$, $\beta=0.25$ & $F\:\searrow$ $[0,100]$    &  $E\:\searrow$ & $R$ $\nearrow$ \\ \cline{2-6}
& $P=10^{-6}$  & $\langle k \rangle=10$, $\beta=0.25$, $\tau_c\:\searrow$ $[0,20)$ &  $F\in[0,100]$  &  $ E\:\nearrow$ &  $ R\:\nearrow$\\ \cline{2-6}
& $P=10^{-6}$  & $\langle k \rangle=10$, $\beta=0.25$, $\tau_c\in[20,60)$ &  $F\in[0,100]$  &  $ E\:\downarrow$ &  $ R\:\downarrow$\\ \cline{2-6}
& $P=10^{-6}$  & $\langle k \rangle=10$, $\beta=0.25$, $\tau_c\:\searrow$ $[60,80]$ &  $F\in[0,100]$  &  $ E\:\nearrow$ &  $ R\:\nearrow$\\ \cline{2-6}
& $P=10^{-6}$   &  $\tau_c=3$,  $\beta=0.25$, $\langle k \rangle\:\nearrow$  $[2,20]$  &  $F\in[0,100]$   &  $E\:\nearrow$ & $R\:\nearrow$  \\ \cline{2-6}
& $P=10^{-6}$   &  $\tau_c=3$,  $\beta=0.25$, $\langle k \rangle\:\searrow$  $[2,20]$  &  $F\in[0,100]$   &  $E\:\searrow$ & $R\:\searrow$  \\ \cline{2-6}
& $P=10^{-6}$   &  $\tau_c=3$,  $\beta=0.25$, $\langle k \rangle\in[2,5]$  &  $F\:\searrow$ $[0,100]$   &   $ E\:\nearrow$ & $ R\:\nearrow$ \\ \cline{2-6}
& $P=10^{-6}$   &  $\tau_c=3$,  $\beta=0.25$, $\langle k \rangle\in[2,5]$  &  $F\:\nearrow$ $[0,100]$   &   $ E\:\searrow$ & $ R\:\searrow$ \\ \cline{2-6}
& $P=10^{-6}$   &  $\tau_c=3$, $\langle k \rangle=5$, $\beta=1$  &  $F\in[0,100]$   &   $E\:\uparrow$ & $  R\:\uparrow$ \\ \cline{2-6}
\hline
\end{tabular}
\end{center}
\end{minipage}
\end{table*}
\subsection{Combined effects of $F$ and $P$}
From many previous research works, it is well established that the strength of the coupling between oscillators (neurons included) is crucial for their synchronization. Essentially, if the coupling strength is zero or below a non-zero threshold, the oscillators cannot synchronize or achieve a certain degree of synchronization. Thus, for a better understanding of synchronization as a function of the STDP parameter $P$, which controls the modification of the synaptic coupling strengths and $F$, it is necessary to first investigate how the average synaptic weight $G$ given in Eq. \eqref{eq:9} varies with $P$ and $F$.
\begin{equation}\label{eq:9}
G = \displaystyle{\Bigg \langle\frac{1}{N^2}\sum\limits_{i=1}^{N}\sum\limits_{j=1}^{N}g_{ij}(t)\Bigg \rangle_t},
\end{equation}
where $\langle\cdot\rangle_t$ is the average over time, $g_{ij}(t)\in[0.001,0.5]$, and $g_{ij}(t=0) \sim \mathcal{N}(0.35,\,0.01)$.

\begin{figure}
\centering
\includegraphics[width=6.0cm,height=4.0cm]{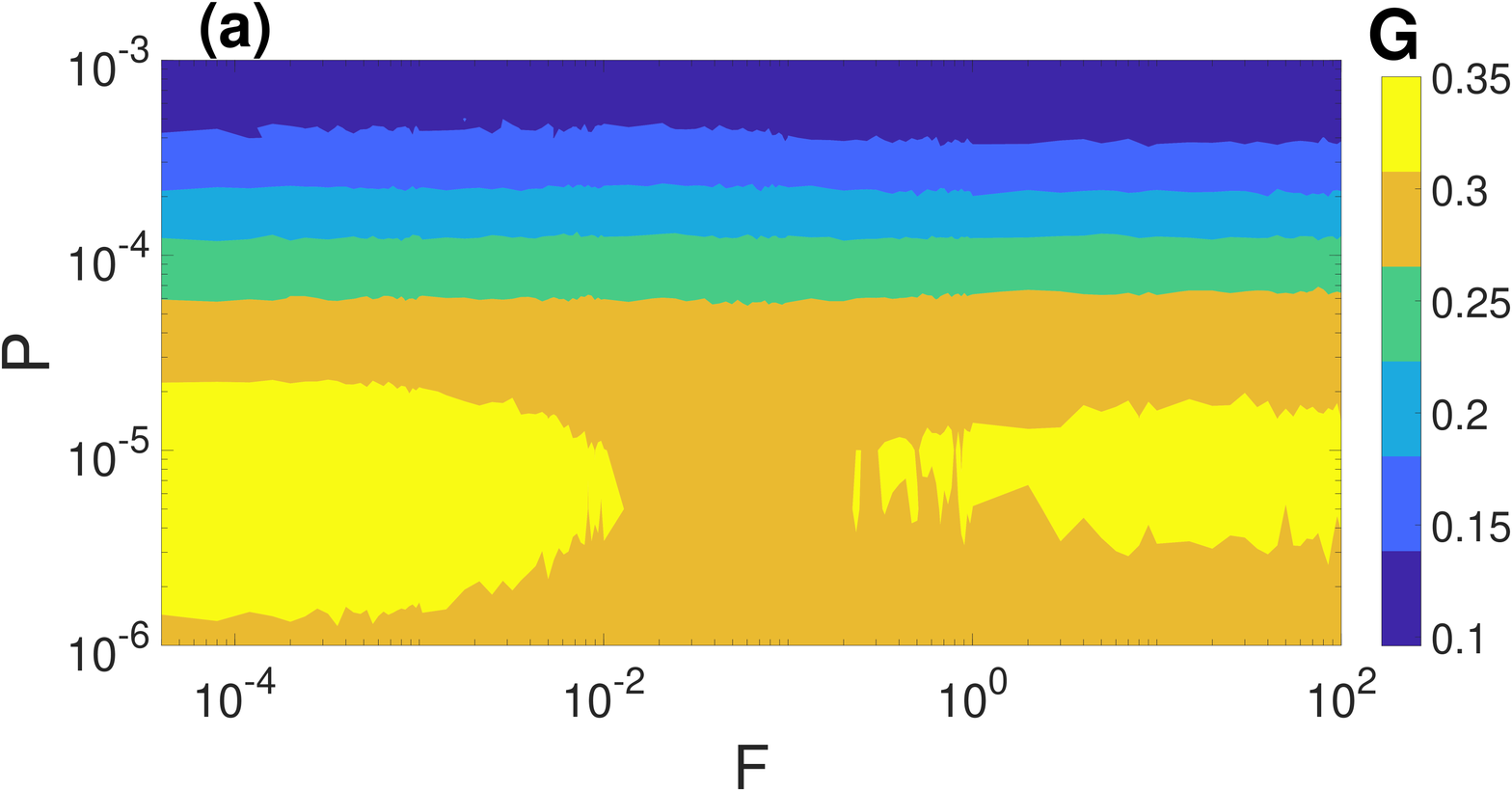}\includegraphics[width=6.0cm,height=4.0cm]{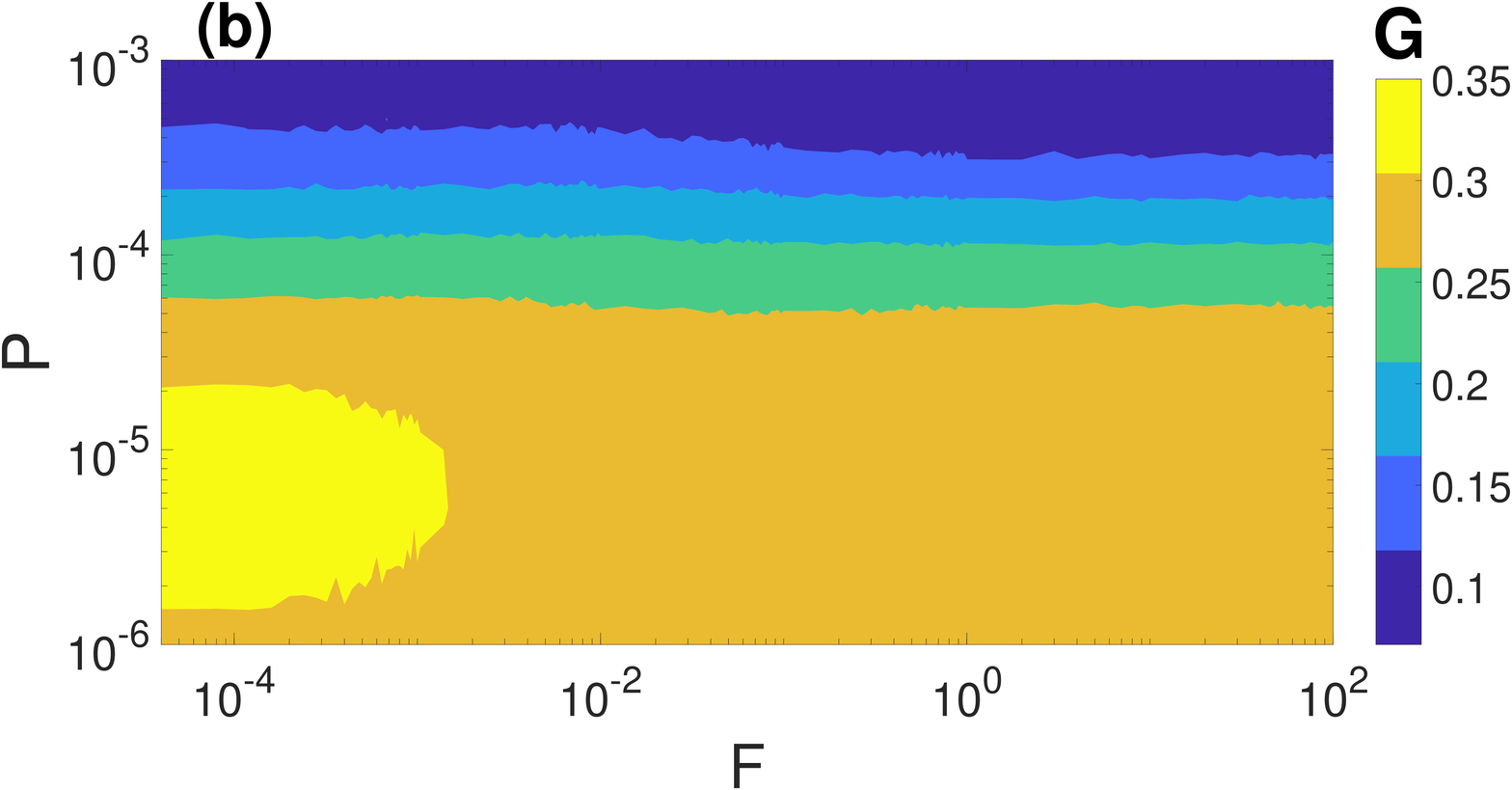}
\caption{Variation of the average synaptic weight $G$  as a function of $P$ and $F$ in \textbf{(a)} small-world ($\beta=0.25$) and \textbf{(b)} random  ($\beta=1$) network. In both topologies, decreasing $P$ strengthens the average synaptic weight after STDP learning, while $F$ has no significant effect on $G$, especially at larger $P$. Parameter values: $\langle k \rangle = 10$, $\tau_c=0.0$, $N=100$.}
\label{fig:1}
\end{figure}

In Figs. \ref{fig:1}\textbf{(a)} and \textbf{(b)} , we present the variation of $G$ as a function of $P$ and $F$ in Watts-Strogatz small-world ($\beta=0.25$) and completely random ($\beta=1$) networks, respectively. We observe that increasing $P$ weakens the average synaptic weight in both small-world and random networks, and at the same time, for a given value of $P$, increasing $F$ has no significant effect on the average synaptic weight. One major difference between the two topologies is that the weakening of synapses after STDP is significantly stronger in the random network, with average synaptic weight reaching a value as low as $G=0.0718$ compared to $G=0.102$ in the small-world network.

The fact that the synapses strengthen with decreasing $P$ leads to the dominant depression of the synaptic weights (as $D/P$ increases and $G$ never exceeds the mean value of the initial synaptic weights distribution $\mathcal{N}(0.35,\,0.01)$) is in agreement with experimental studies \cite{bi1998synaptic,feldman2005map}. Hence, we expect that decreasing $P$ would favor synchronization.

\marius{The variation of $G$ in Fig. \ref{fig:1} is robust, as extensive numerical simulations (not shown) indicate that $G$ displays the same qualitative behavior with respect to $P$ and $F$ and for other values of the synaptic time delay $\tau_c\in[0,80]$, average degree connectivity  $\langle k \rangle\in[2,30]$, rewiring probability $\beta\in(0,1]$, and network size $N\in[80,120]$. Thus, even though $G$ is not a variable main interest in our study, it is worth pointing out that the way the dynamics of $G$ relate to the degree of CS and PS can be inferred from its inverse and monotonic variation with $P$. In the next subsections, we will investigate the combined effect of $F$ and a network parameter ($\tau_c$,  $\langle k \rangle$, $\beta$) on synchronization at the smallest value of $P$ $(=1.0\times10^{-6})$, i.e., the largest value of average synaptic strength $G(=0.35)$ in the network.} 

\begin{figure}
\centering
\includegraphics[width=6.0cm,height=4.0cm]{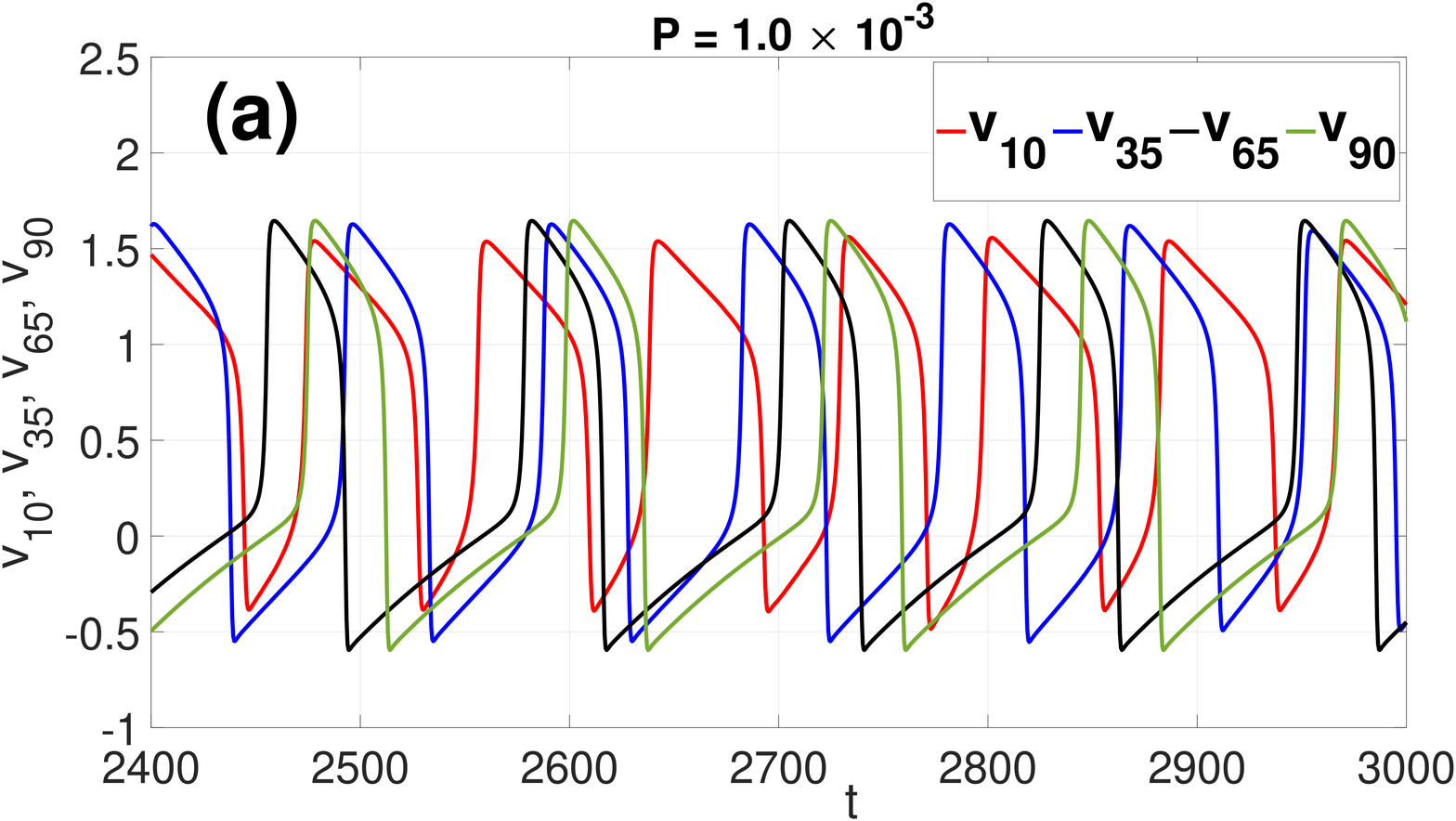}\includegraphics[width=6.0cm,height=4.0cm]{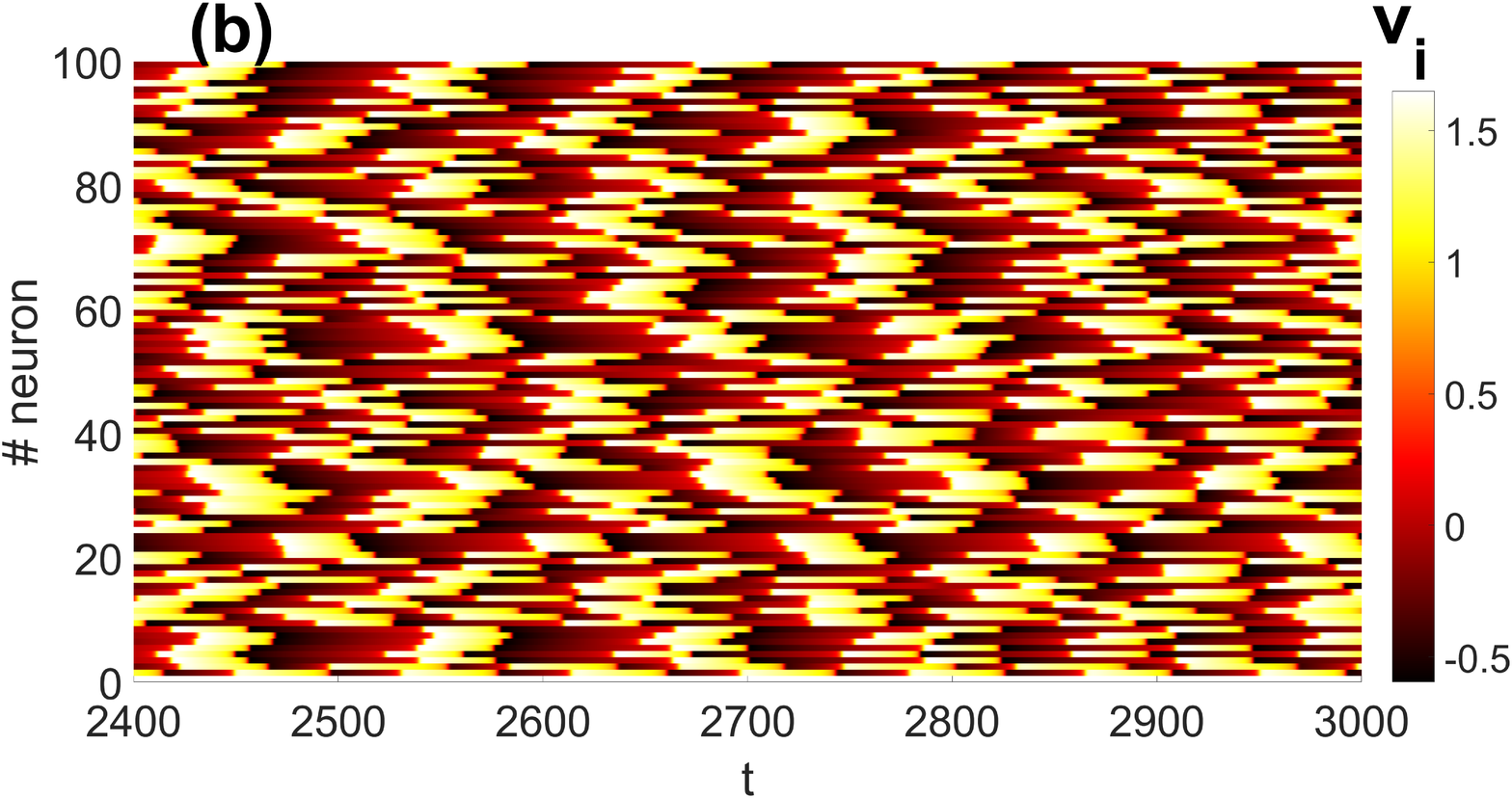}
\includegraphics[width=6.0cm,height=4.0cm]{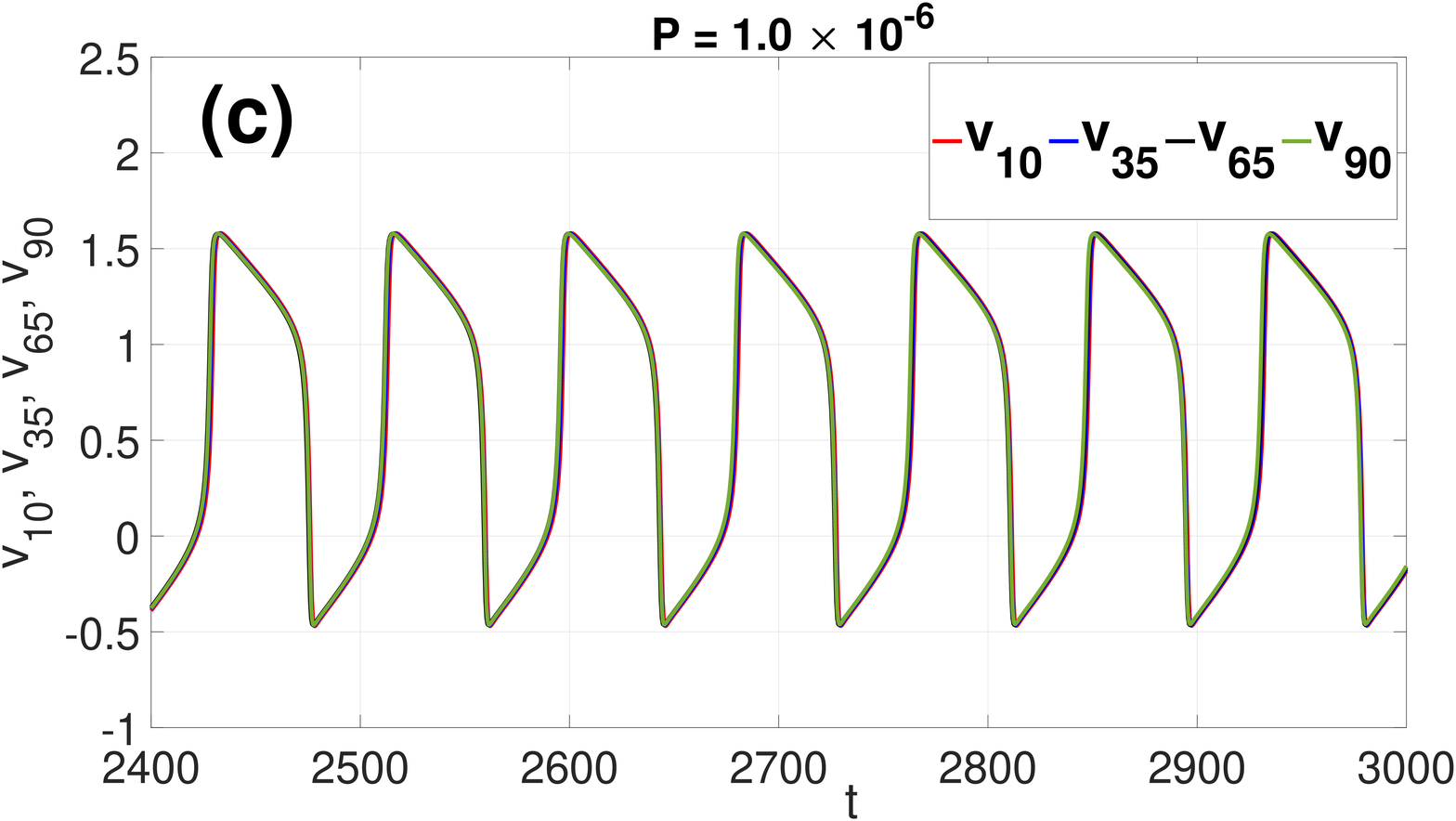}\includegraphics[width=6.0cm,height=4.0cm]{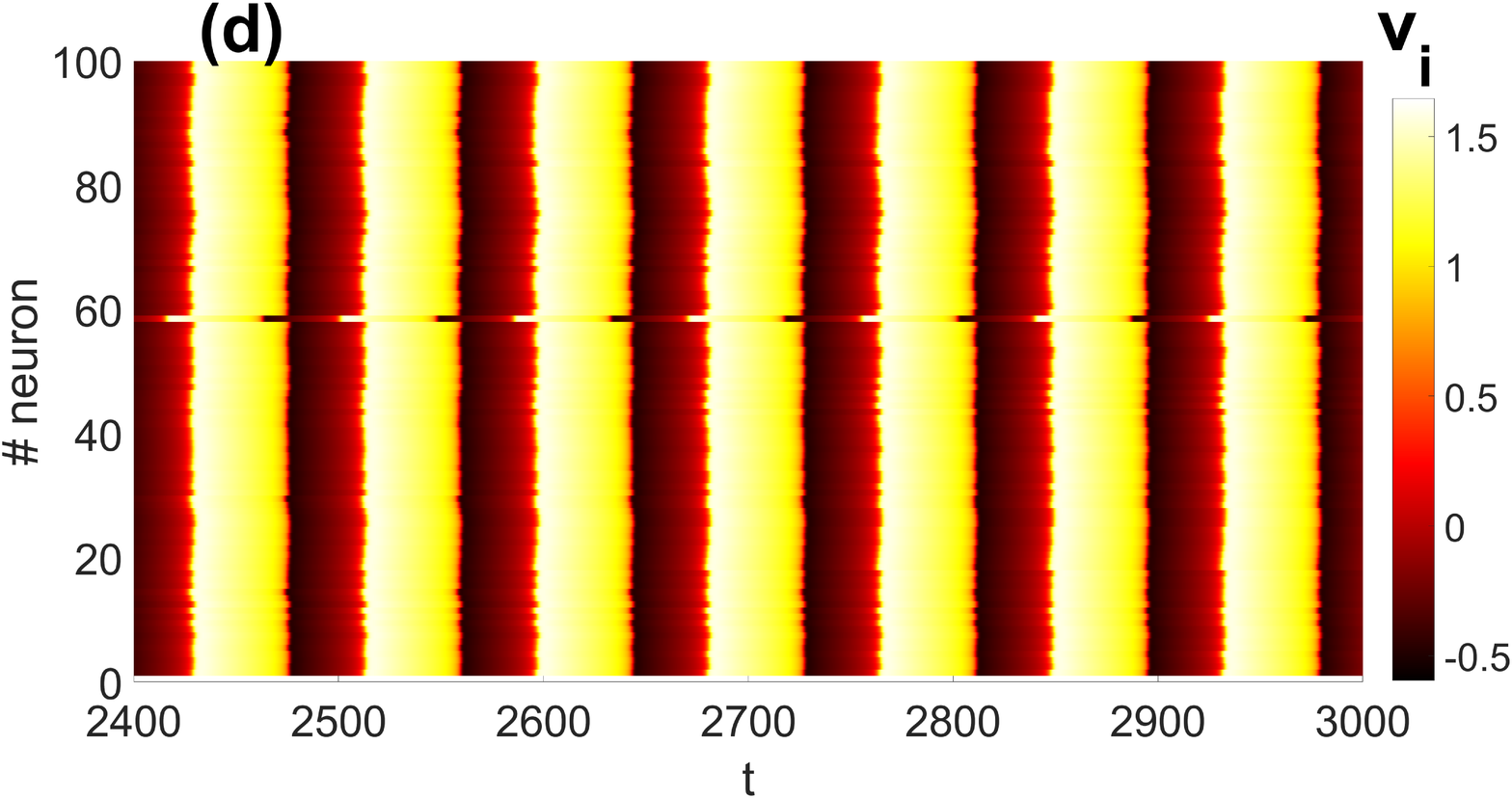}
\caption{\marius{\marius{Time series of some neurons' membrane potential in \textbf{(a)} and corresponding spatiotemporal pattern in \textbf{(b)} 
of a small-world network ($\beta=0.25$) with $P=1.0\times10^{-3}$ exhibiting poor degree CS and PS.
Time series of some neurons' membrane potential in \textbf{(c)} and corresponding spatiotemporal pattern in \textbf{(d)} 
of the random network ($\beta=1$) with $P=1.0\times10^{-6}$ exhibiting a high degree of CS and PS.
Parameter values: $F=100$, $\langle k \rangle = 10$, $\tau_c=0.0$, $N=100$.}}}
\label{fig:11}
\end{figure}

\begin{figure}
\centering
\includegraphics[width=6.0cm,height=4.0cm]{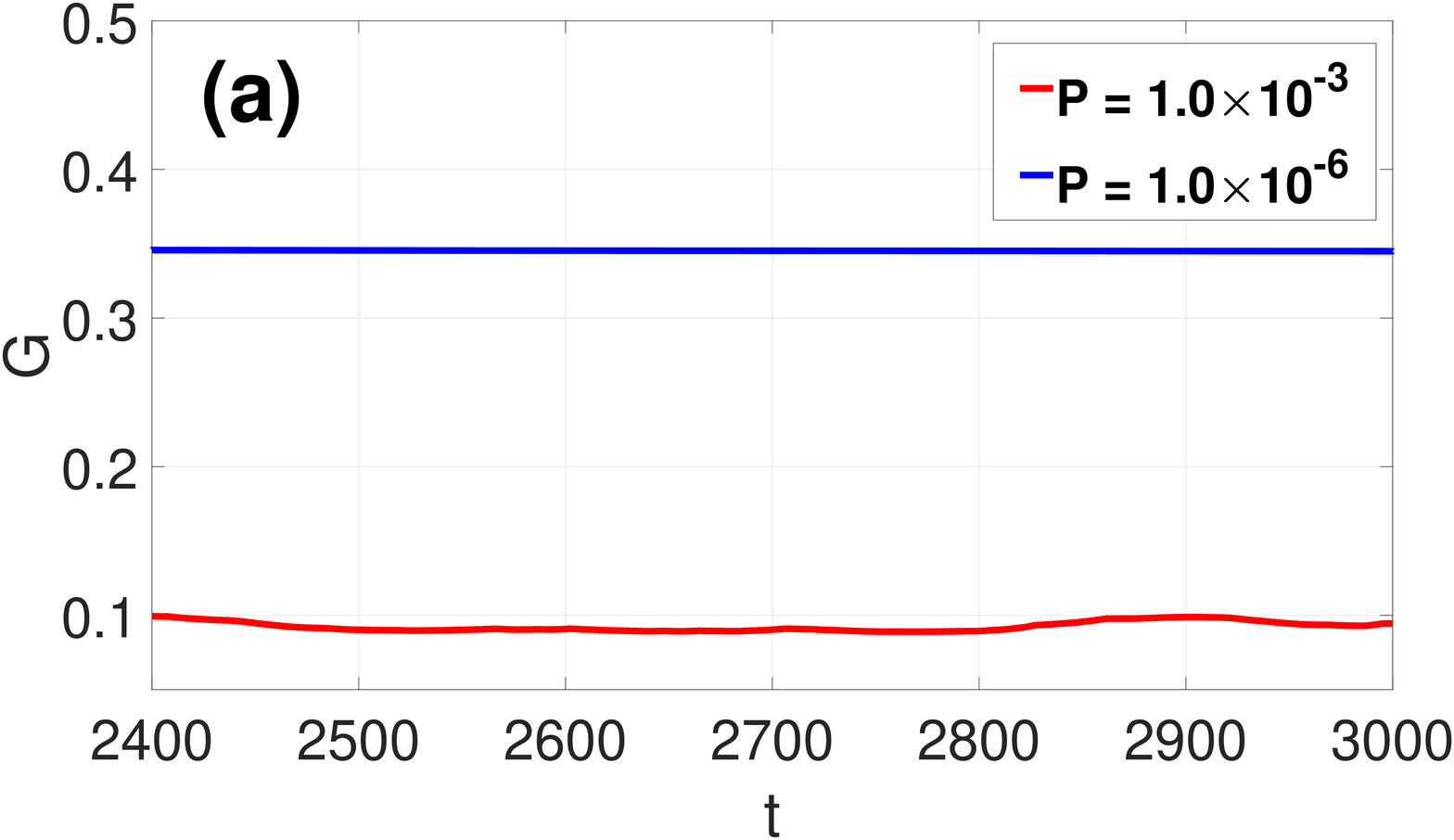}\includegraphics[width=6.0cm,height=4.0cm]{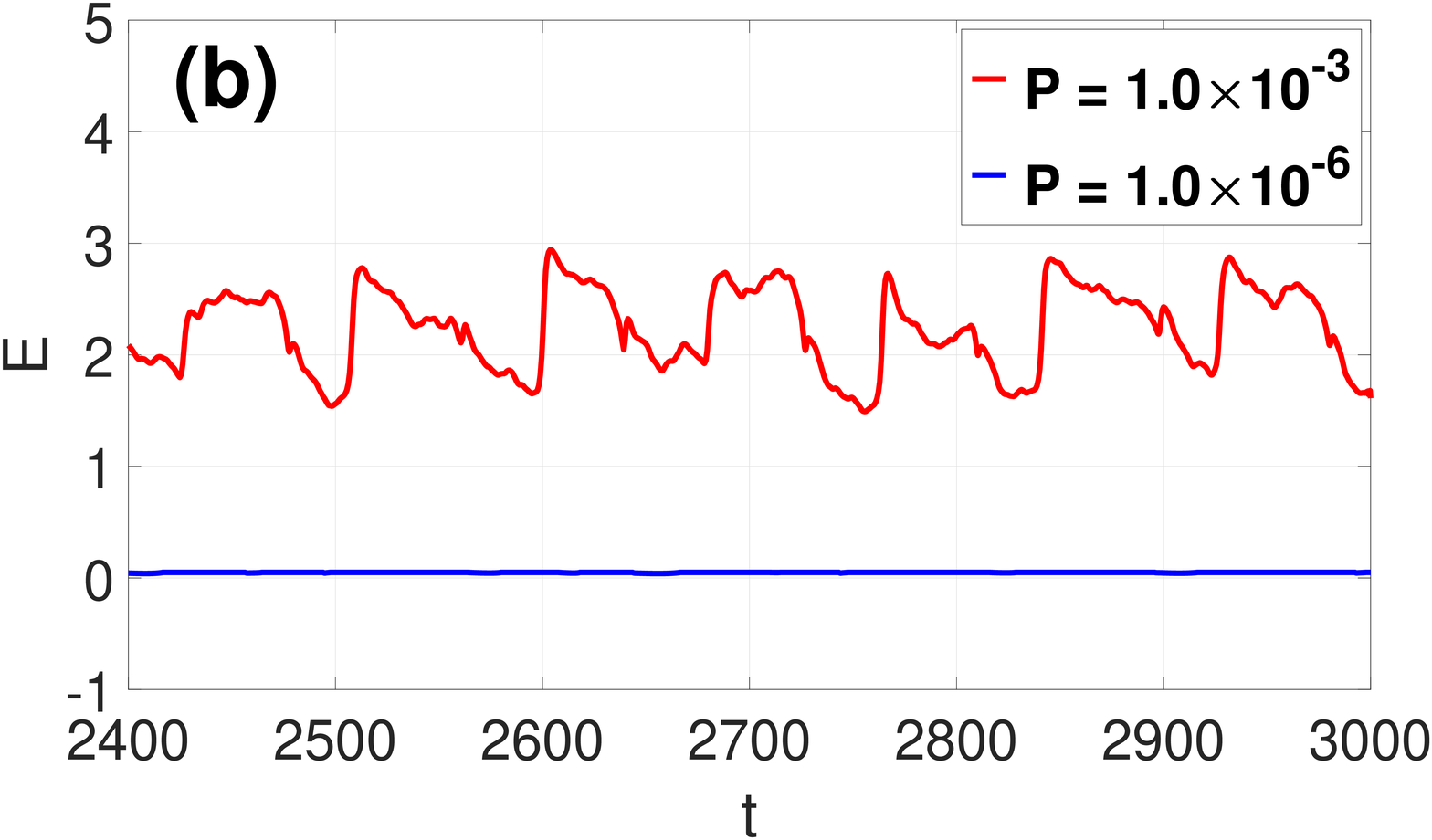}
\includegraphics[width=6.0cm,height=4.0cm]{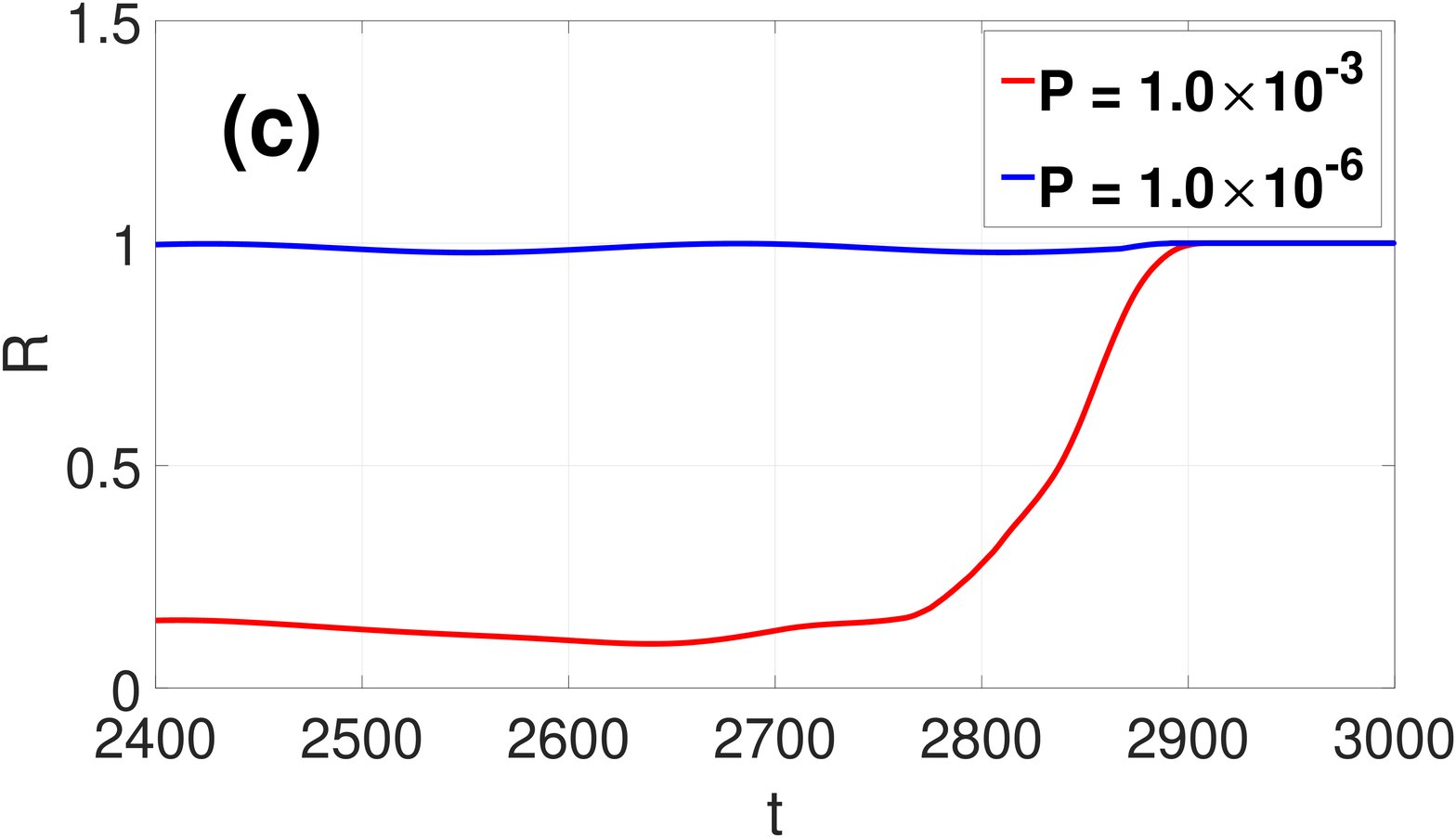}
\caption{\marius{
Time series of average synaptic weight $G$ in \textbf{(a)} for small-world ($\beta=0.25$, red curve) and random ($\beta=1.0$, blue curve) networks at various STDP parameter $P$ values. Time series of average error of membrane potential traces in \textbf{(b)} for small-world ($\beta=0.25$, red curve) and random ($\beta=1.0$, blue curve) networks at different $P$. Time series of average error for Kuramoto order parameter $R$ in \textbf{(a)}  for small-world ($\beta=0.25$, red) and random ($\beta=1.0$, blue) networks at different $P$. Parameter values: $F=100$, $\langle k \rangle = 10$, $\tau_c=0.0$, $N=100$.
}}
\label{fig:111}
\end{figure}

\marius{In Figs. \ref{fig:11}\textbf{(a)} and \textbf{(b)}, we show, respectively, the time series of a few spiking neurons and the spatiotemporal pattern of all the spiking neurons in a small world network of Fig. \ref{fig:1}\textbf{(a)}, when the STDP parameter is relatively large, i.e., $P=1.0\times10^{-3}$, leading to a weak average synaptic strength $G\approx0.1$. In these figures, it can be seen the neurons exhibit a poor degree of CS (see the red curve in Fig. \ref{fig:111}\textbf{(b)}) and a poor degree of PS at early times of the time-series (see the red curve in Fig. \ref{fig:111}\textbf{(c)}) due to the weak average synaptic strength (see the red curve in Fig. \ref{fig:111}\textbf{(a)}). Figures \ref{fig:11}\textbf{(c)} and \textbf{(d)} display the time series of a few spiking neurons and the spatiotemporal pattern of all spiking neurons in the random network of Fig. \ref{fig:1}\textbf{(b)}, when the STDP parameter is relatively small, i.e., $P=1.0\times10^{-6}$, leading to a stronger average synaptic strength $G\approx0.35$. In this case, the neurons exhibit good degree of CS (see the blue curve in Fig. \ref{fig:111}\textbf{(b)}) and a good degree of PS (see the blue curve in Fig. \ref{fig:111}\textbf{(c)}) due to the stronger average synaptic strength (see the blue curve in Fig. \ref{fig:111}\textbf{(a)}).}

\marius{The red curve in Fig. \ref{fig:111}\textbf{(a)} represents the time series of the averaged synaptic weight $G$ of the small-world network ($\beta=0.25$) when the STDP parameter is relatively large $P=1.0\times10^{-3}$ --- just as in Figs. \ref{fig:11}\textbf{(a)} and \textbf{(b)}. In this case, we can see that $G$ saturates at a relatively low value. Hence, the poor degree of CS, as indicated in Figs. \ref{fig:11}\textbf{(a)} and \textbf{(b)}, and the relatively high synchronization error $E$ represented by the red curve in Figs. \ref{fig:111}\textbf{(b)}. Furthermore, for this same value of $P(=1.0\times10^{-3})$, we also observe a poor degree of PS measured by the relatively low Kuramoto order parameter $R$ represented by the red curve in Fig. \ref{fig:111}\textbf{(c)}. 
}

\marius{
However, towards the end of the time series in Fig. \ref{fig:111}\textbf{(b)}, the red curve increases from a relatively low value to higher values near 1, indicating a better degree of PS. This explains why in Figs. \ref{fig:11}\textbf{(a)}, some neurons towards the end of the time series turn to synchronize their spiking times, leading to a higher degree of PS. Nevertheless, it is worth noting that CS is still very poor as most neurons have synchronized only their spiking times and not the traces of their membrane potentials.
}

\marius{Furthermore, we observe that towards the end of the time series in Fig. \ref{fig:111}\textbf{(a)}, there is no growth in the average synaptic strength $G$. Hence, the synaptic strength is not responsible for this improvement in the degree of PS toward the end of the time series in Fig. \ref{fig:11}\textbf{(a)} and the red curve in Fig. \ref{fig:111}\textbf{(c)}. This behavior is explained by the fact that our oscillators (FHN neurons in Eq. \eqref{eq:1}) are identical. Thus, with a sufficiently long transient time, identical oscillators with weak coupling can still synchronize because of the similarity of their attractors in phase space.  In this case, the oscillators adjust their phases to align in a specific relationship, while their amplitudes may differ (hence, the poor degree of CS). 
This PS occurs due to the shared properties of the oscillators, such as having identical parameter values, natural frequencies, and similar dynamical behaviors. When the coupling between the identical oscillators weakens, their interaction is not strong enough to force CS. However, some or most oscillators may occasionally achieve a state of PS where their phases become correlated -- like at the end of the time series in Fig. \ref{fig:11}\textbf{(a)} (where the last spiking time of some neuron coincide), leading to a higher degree of PS as indicated by the higher values of $R$ toward the end of the time series in Fig.\ref{fig:111}\textbf{(c)}.
}

\marius{
On the other hand, the blue curve in Fig. \ref{fig:111}\textbf{(a)} represents the time series of the averaged synaptic weight $G$ of the random network ($\beta=1$) when the STDP parameter is relatively small $P=1.0\times10^{-6}$ --- just as in Figs. \ref{fig:11}\textbf{(c)} and \textbf{(d)}. In this case, we can see that $G$ saturates at a relatively high value. Hence, the high degree of CS, as indicated in Figs. \ref{fig:11}\textbf{(c)} and \textbf{(d)}, and the relatively low synchronization error $E$ represented by the blue curve in Fig. \ref{fig:111}\textbf{(b)}. Furthermore, for this same value of $P(=1.0\times10^{-6})$, we also observe a high degree of PS measured by the relatively high Kuramoto order parameter $R$ represented by the blue curve in Fig. \ref{fig:111}\textbf{(c)}.}

\begin{figure*}
\centering
\includegraphics[width=6.0cm,height=4.0cm]{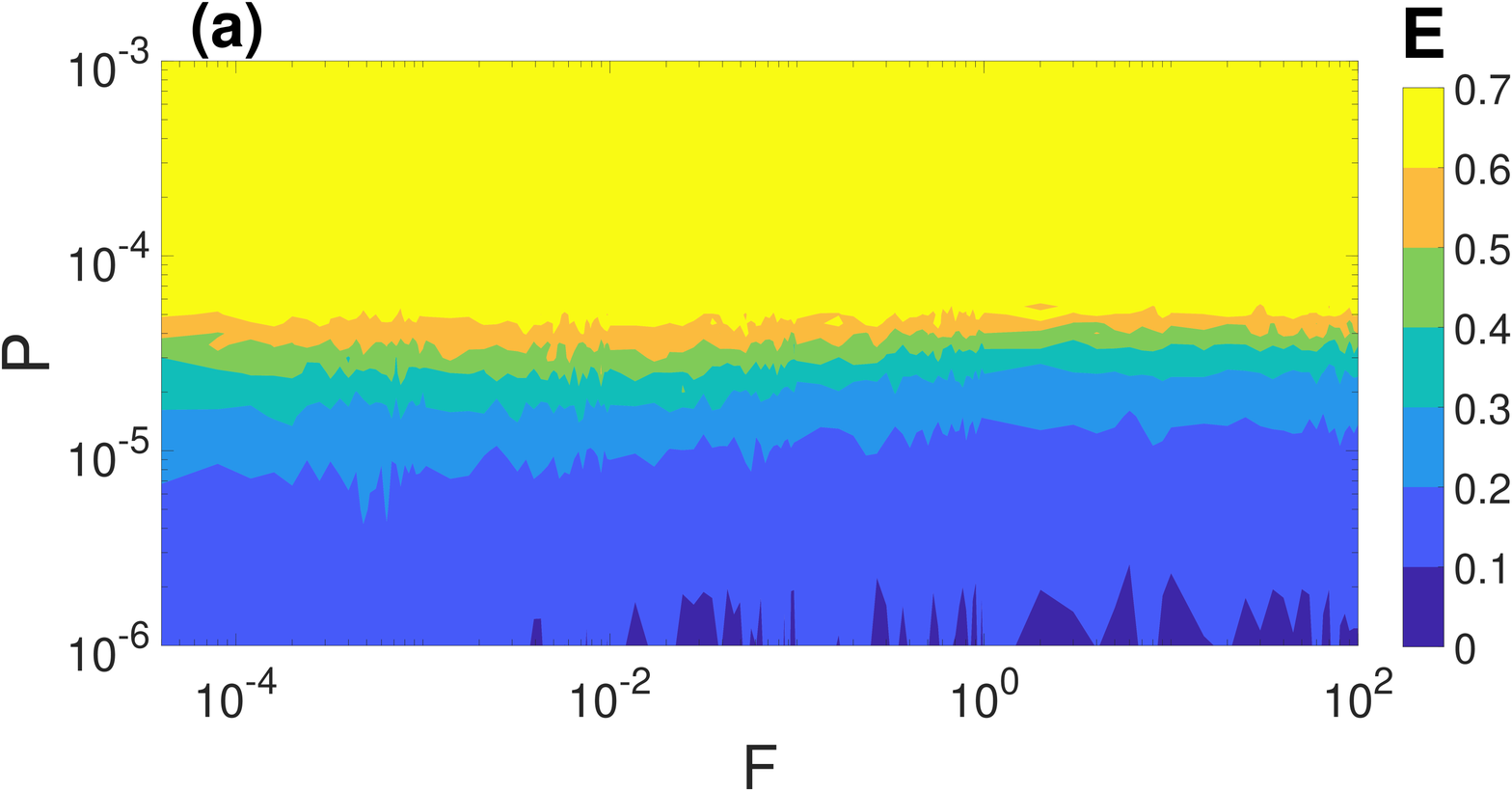}
\includegraphics[width=6.0cm,height=4.0cm]{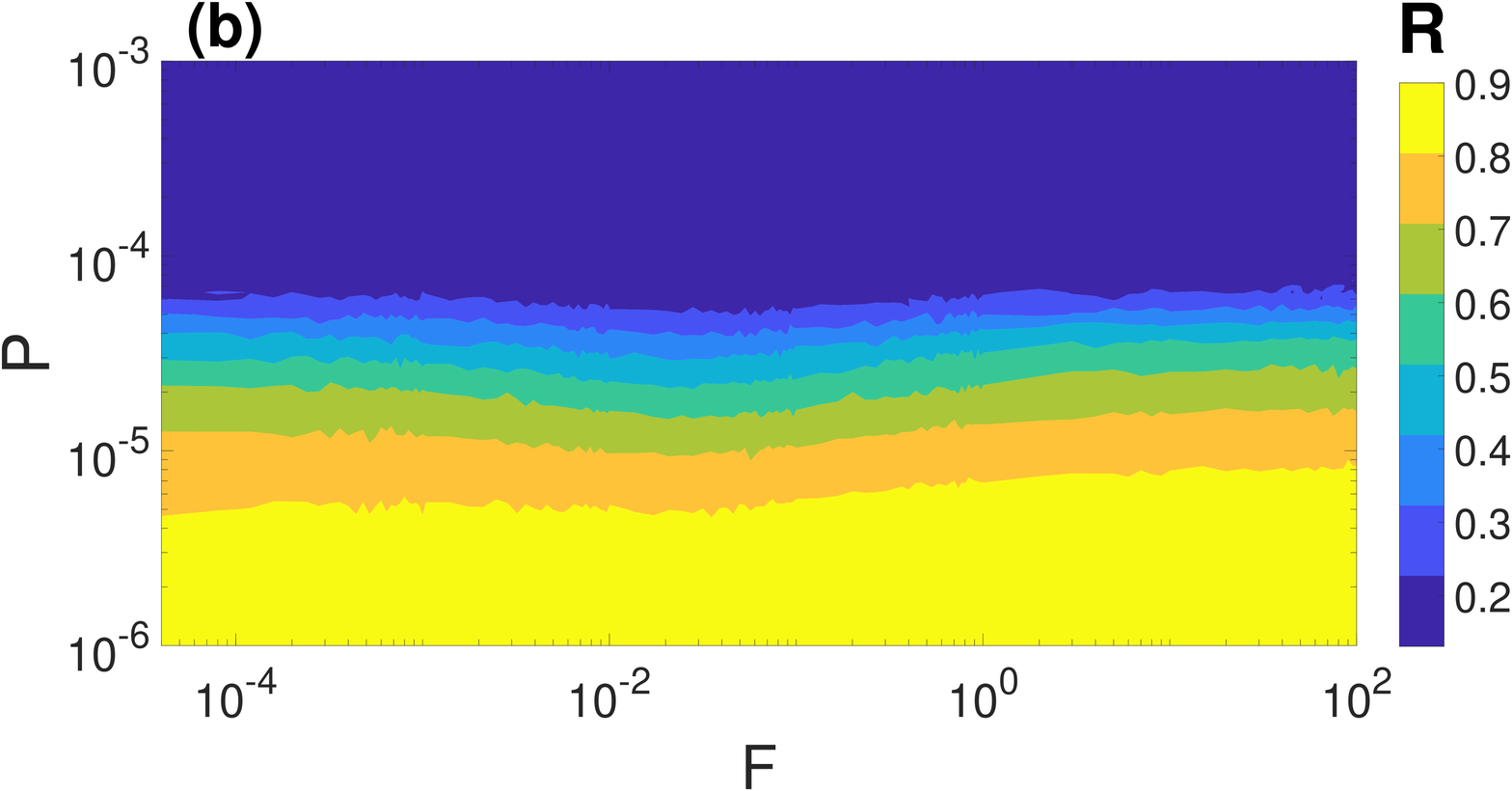}\\
\includegraphics[width=6.0cm,height=4.0cm]{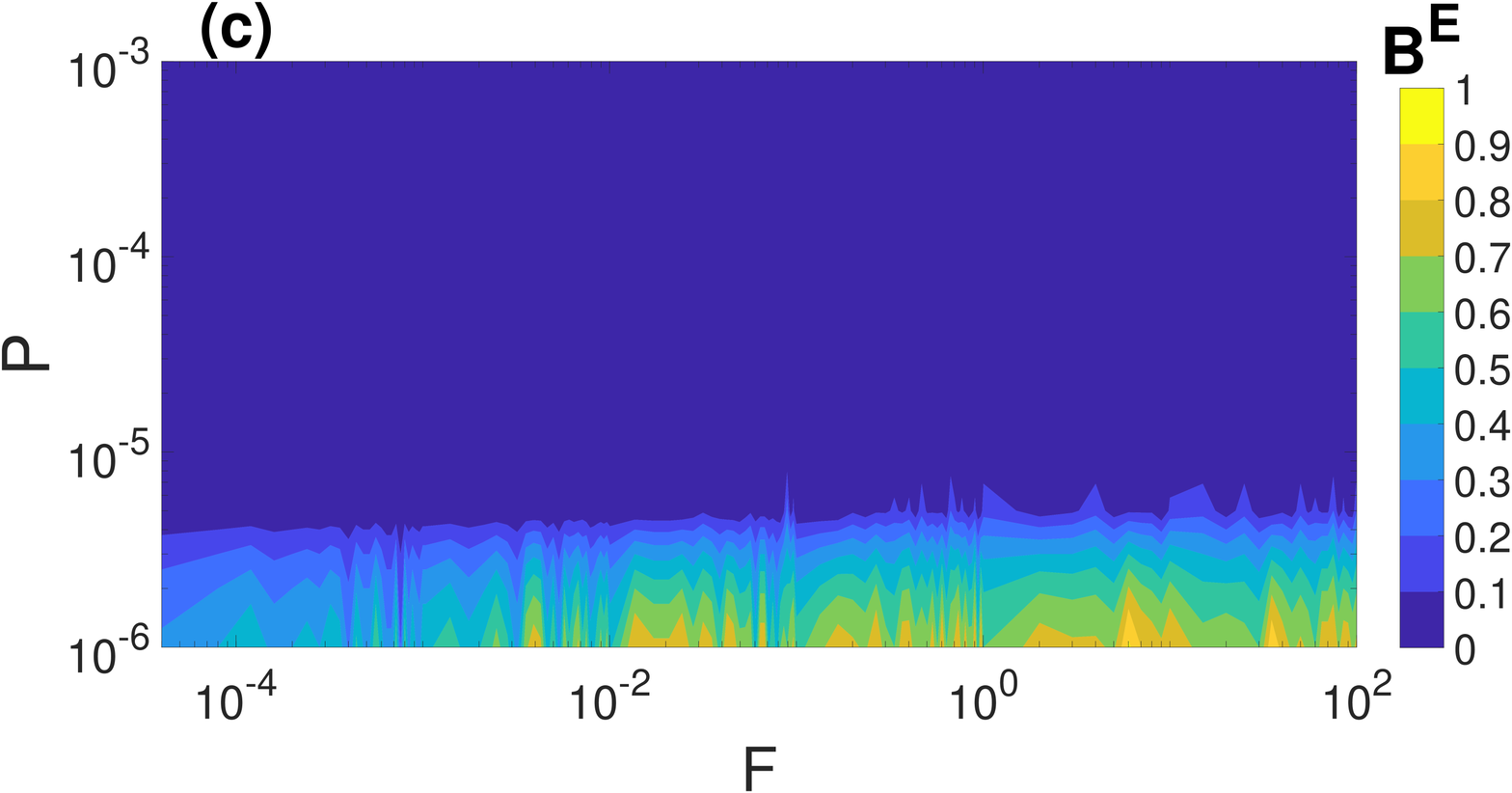}
\includegraphics[width=6.0cm,height=4.0cm]{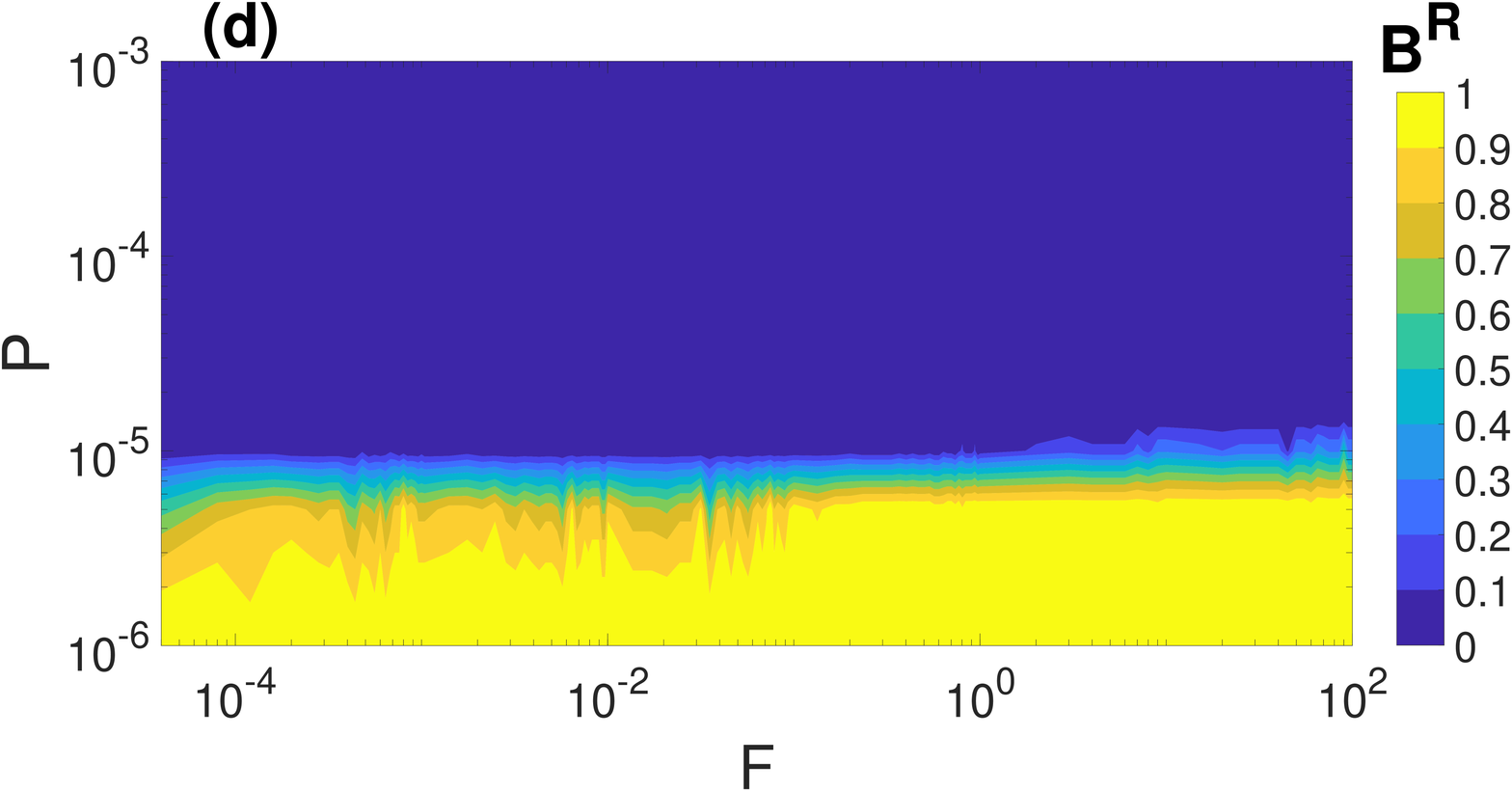}
\caption{Variation in the degree of synchronization and the corresponding global stability w.r.t. $P$ and $F$ in a small-world network.  \textbf{(a)} and \textbf{(c)}: degree of CS and the corresponding basin stability measure. \textbf{(b)} and \textbf{(d)}: degree of PS and the corresponding basin stability measure. Parameter values: $\langle k \rangle = 10$,  $\beta= 0.25$,  $\tau_c=0.0$, $N=100$.}
\label{fig:2}
\end{figure*}

\marius{
In the rest of the paper, we present the behaviors of CS and PS in the small-world and random networks as a function of each network parameter and the network rewriting frequency $F$, at the best STDP parameter value (i.e., $P=1.0\times10^{-6}$) for both types of synchronization.} In Figs. \ref{fig:2}\textbf{(a)} and \textbf{(b)}, we depict the variations in the degree of CS and PS as a function of $P$ and $F$ in a small-world network ($\beta=0.25$), respectively. It is evident from these two figures that decreasing the value of $P$ (i.e., strengthening the average synaptic weights in the network after STDP, as shown in  Fig. \ref{fig:1}) enhances the degree of CS (i.e., $E \to 0$) and the degree of PS (i.e., $R \to 1$). At the same time, for any given value of $P$, increasing the value of $F$ has no significant effect on the degree of CS and PS, except in the case of CS for very small values of $P$ $(\approx10^{-6})$, where increasing $F$ occasionally enhances the degree of CS to almost full synchrony ($E\approx0$). This implies that when the average synaptic weight is strong, a more rapidly changing small-world network can achieve larger windows of CS. Comparing the degree of CS and PS, we observe that a relatively weaker average synaptic weight (controlled by $P$) is required to achieve a high degree of PS (shown in light yellow) as opposed to CS, which requires a much stronger average synaptic weight to attain a high degree.

In Figs. \ref{fig:2}\textbf{(c)} and \textbf{(d)}, we present the basin stability of CS and PS corresponding to Figs. \ref{fig:2}\textbf{(a)} and \textbf{(b)}, respectively. Figure \ref{fig:2}\textbf{(c)} indicates the highest degrees of CS (i.e., the dark blue regions in Fig. \ref{fig:2}\textbf{(a)}, with $P\approx10^{-6}$ and $E<10^{-1}$) that are not globally stable (i.e., $B^{E}<1$) in the prescribed region of phase space. Instead, we have the co-existence of a desynchronized state and a synchronized state (the latter being more probable than the former since $0.7<B^{E}<1$). Furthermore, it can be observed in Fig. \ref{fig:2}\textbf{(c)} that when $P\approx10^{-6}$, increasing $F$ leads to an increase in $B^E$, indicating that small-world network with more rapidly switching synapses and a strong average synaptic weight after STDP will yield a globally stable CS.  Figure \ref{fig:2}\textbf{(d)} indicates that the highest degree PS achieved in Fig. \ref{fig:2}\textbf{(b)} (light yellow regions) is globally stable (i.e., $B^{R}\approx1$) for slightly lower values of $P$. It can also be seen that for $10^{-6}<P<10^{-5}$, increasing $F$ yields an increase in $B^{R}$ from 0.6 to almost 1, indicating that, just like with CS, rapidly switching synapses increases the basin stability of PS. Moreover, comparing the basins stability of CS and PS, it is clear that PS is more stable than CS in the above-prescribed region of phase space.  Qualitatively similar results (not shown) are obtained for the random network ($\beta=1$). In the following sections, when we refer to the optimal value of $P$, we specifically indicate $P=1.0\times10^{-6}$. The results in Fig. \ref{fig:2} indicate that this value of $P$ yields the highest degrees of CS and PS.

\marius{In summary, in the $P-F$ parameter plane, decreasing $P$ (which increases the weakening effect of STDP on the synaptic weights) and increasing $F$ (which speeds up the swapping rate of synapses between neurons) leads to a more stable and higher degree of CS and PS in both the small-world and random networks, provided that $\tau_c$, $\beta$, and $\langle k \rangle$ are fixed at suitable values.}

\subsection{Combined effect of $F$ and $\tau_c$ at the optimal $P$}
\begin{figure*}[h]
\centering
\includegraphics[width=6.0cm,height=4.0cm]{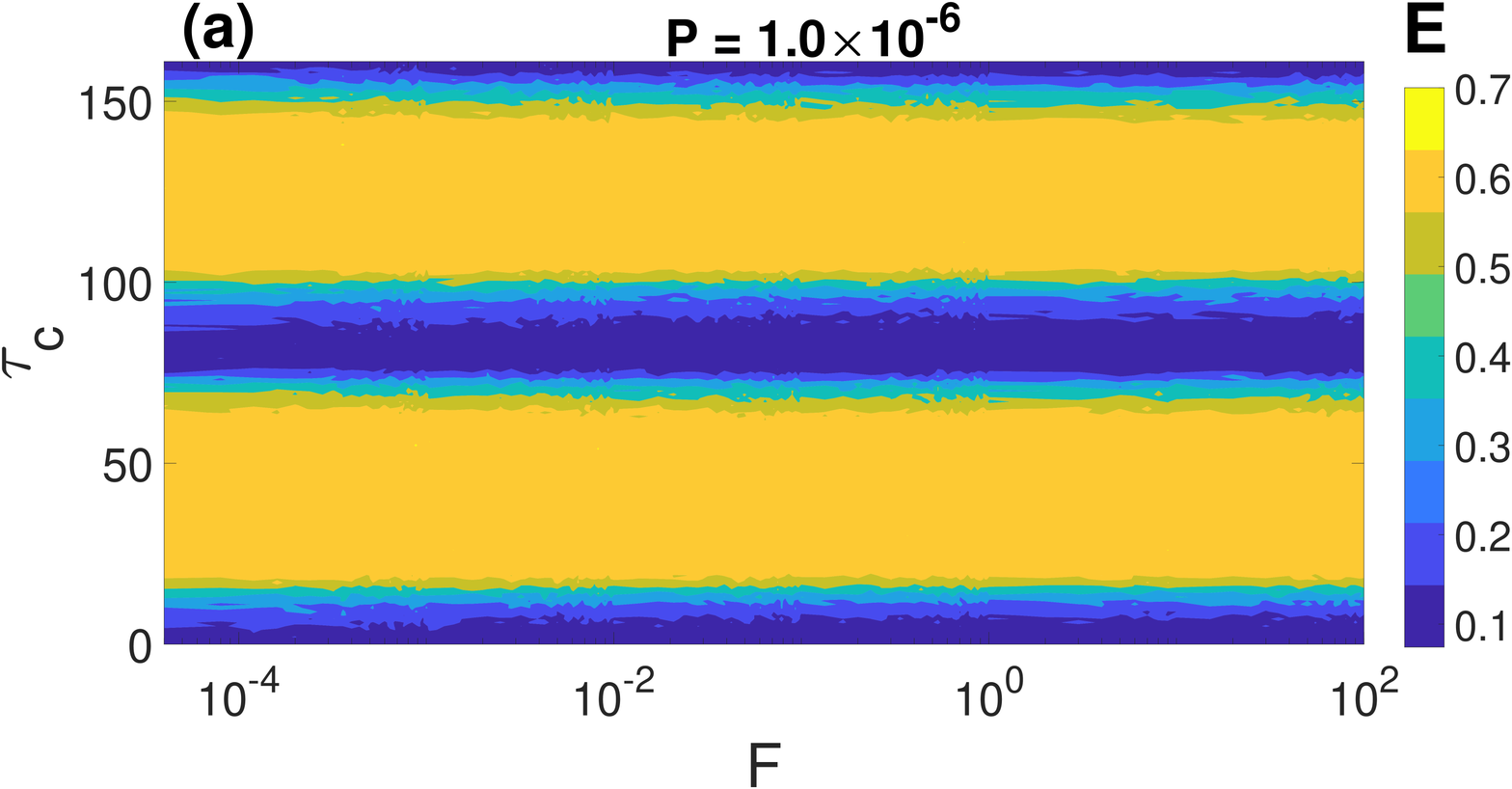}
\includegraphics[width=6.0cm,height=4.0cm]{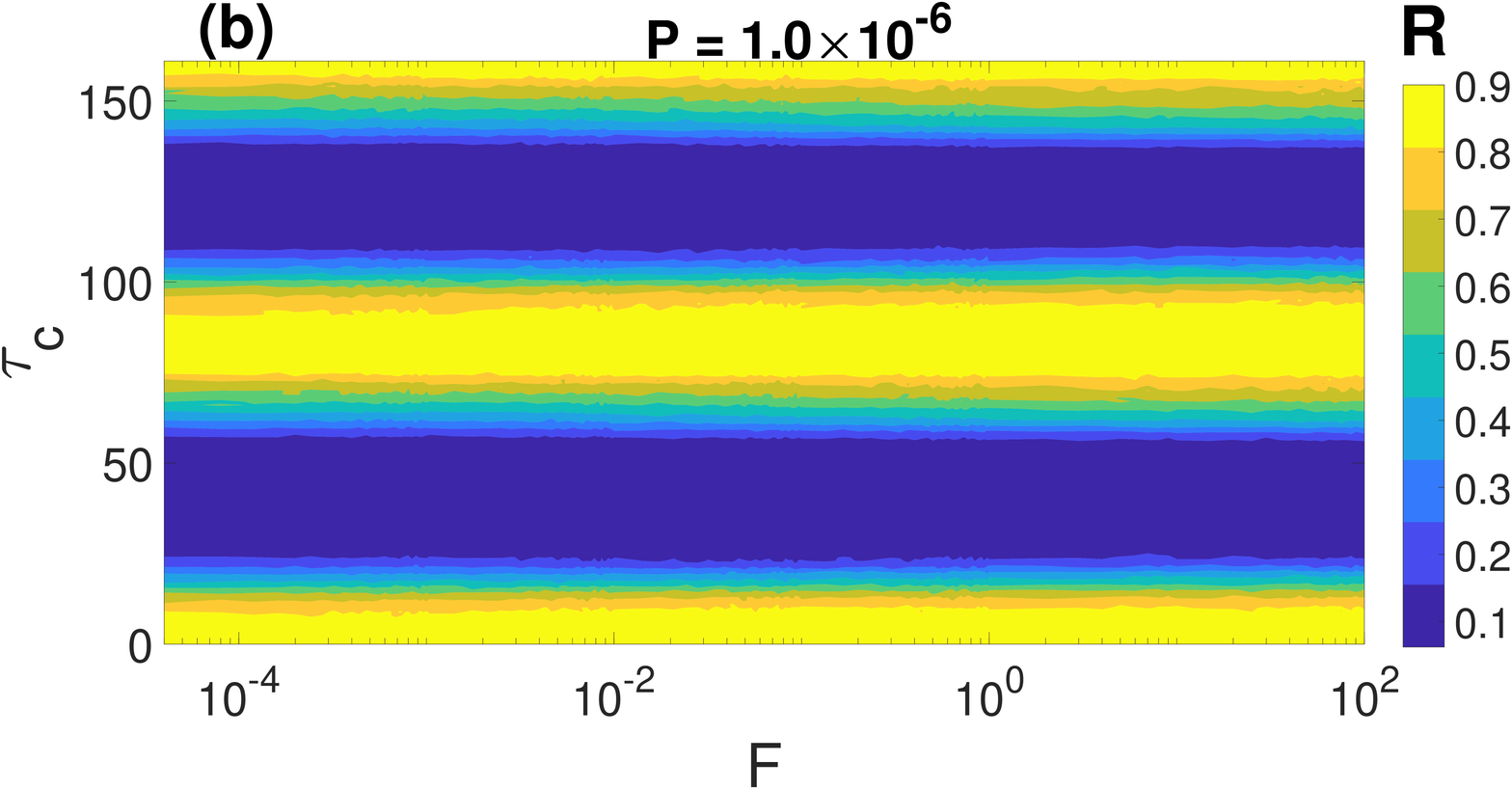}\\
\includegraphics[width=6.0cm,height=4.0cm]{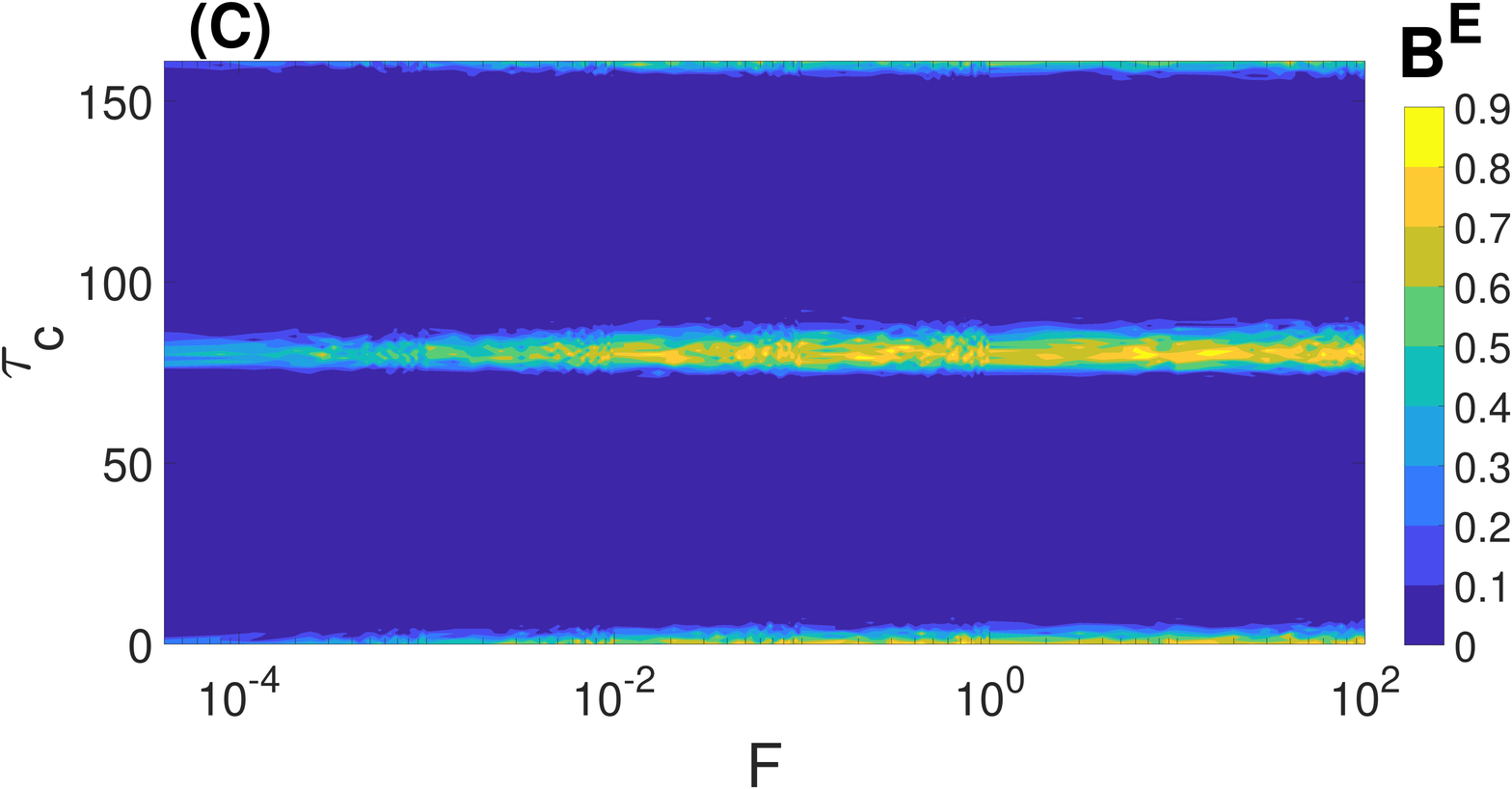}
\includegraphics[width=6.0cm,height=4.0cm]{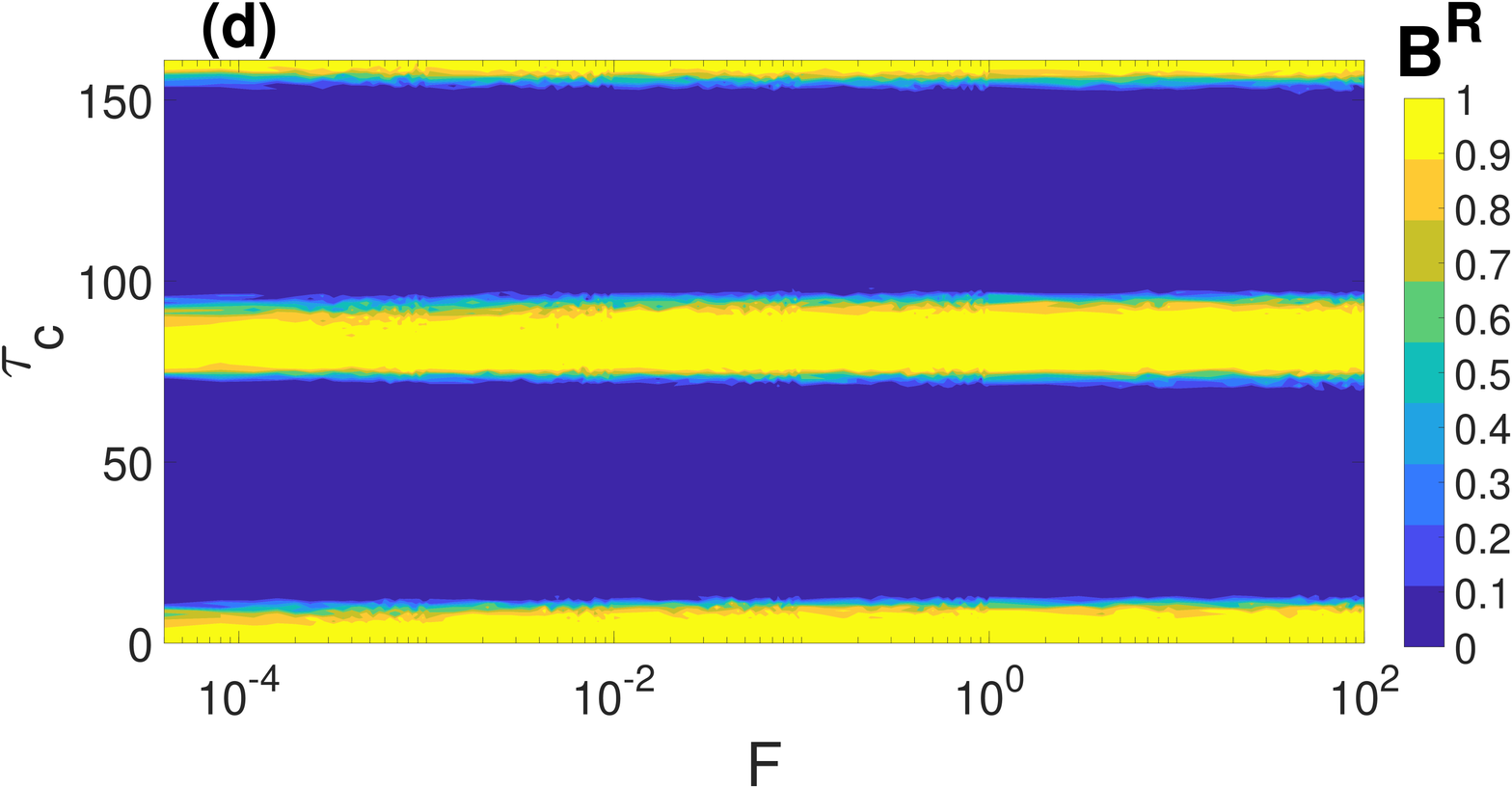}
\caption{Variation in the degree of synchronization and the corresponding global stability w.r.t. $\tau_c$ and $F$ at the optimal value of $P$ indicated. For a small-world network. \textbf{(a)} and \textbf{(c)}: degree of CS and the corresponding basin stability measure. \textbf{(b)} and \textbf{(d)}: degree of PS and the corresponding basin stability measure.  Parameter values: $\langle k \rangle = 10$,  $\beta= 0.25$,  $\tau_c=0.0$, $N=100$.}
\label{fig:3}
\end{figure*}

In Figs. \ref{fig:3}\textbf{(a)} and \textbf{(b)}, we present the variations in the degree of CS and PS as a function of the synaptic time delay $\tau_c\in[0,160]$ and $F$ at the optimal value of $P$ in a small-world network. The results indicate that the small-world network exhibits intermittent CS and PS, irrespective of the switching frequency of synapses $F$. 
Next, we provide a mathematical explanation for intermittent CS and PS as $\tau_c$ increases. First, we recall that if a deterministic delayed differential equation is generally given as $\dot{x}= f(x(t),x(t-\tau_c))$, where $\tau_c$ is the time delay, possesses a solution $x(t)$ with period $\overline{\tau}$, then $x(t)$ also solves $\dot{x}= f(x(t),x(t-\tau_c-n\overline{\tau}))$, for all positive integers $n\in\mathbb{N}$.
It suffices to check if the distance between the horizontal bands of the maximum degree of CS and PS in Figs. \ref{fig:3}\textbf{(a)} and \textbf{(b)}, compares to  the average (over the total number of neurons) interspike interval (ISI), alias period of the neural activity which is computed and given by $ISI\approx80$. It is observed from Fig. \ref{fig:3}\textbf{(a)} that three deep blue horizontal bands where the network exhibits the highest degree of CS are equidistant, and the distance between each is given $\overline{\tau}\approx80\approx ISI$. Hence, the synchronization pattern for CS repeats itself $n$ times after $n\overline{\tau}$, n=0,1,2,..., waiting time. This explanation applies to the case of PS in Fig. \ref{fig:3}\textbf{(b)}. 

Figures \ref{fig:3}\textbf{(c)} and \textbf{(d)} display the basin stability measure of CS and PS presented in Figs. \ref{fig:3}\textbf{(a)} and \textbf{(b)}, respectively. It can be observed from Fig. \ref{fig:3}\textbf{(c)} that higher rewiring frequencies increase the basin stability of CS, especially at intermediate time delays, i.e., at $\tau_c\approx80$. Furthermore, we can again see that the highest degree of CS is less stable than that of PS.  In the case of the random network ($\beta=1$), we have obtained qualitatively similar results (not shown).

\marius{In summary, in the $\tau_c-F$ parameter plane, both small-world and random networks display intermittent CS and PS as $\tau_c$ increases, with the highest degrees of CS and PS occurring when the synaptic time delay $\tau_c$ is multiple of the average inter-spike interval of the networks.}

\subsection{Combined effect of $F$ and $\langle k \rangle$ at the optimal $P$}
\begin{figure*}
\centering
\includegraphics[width=6.0cm,height=4.0cm]{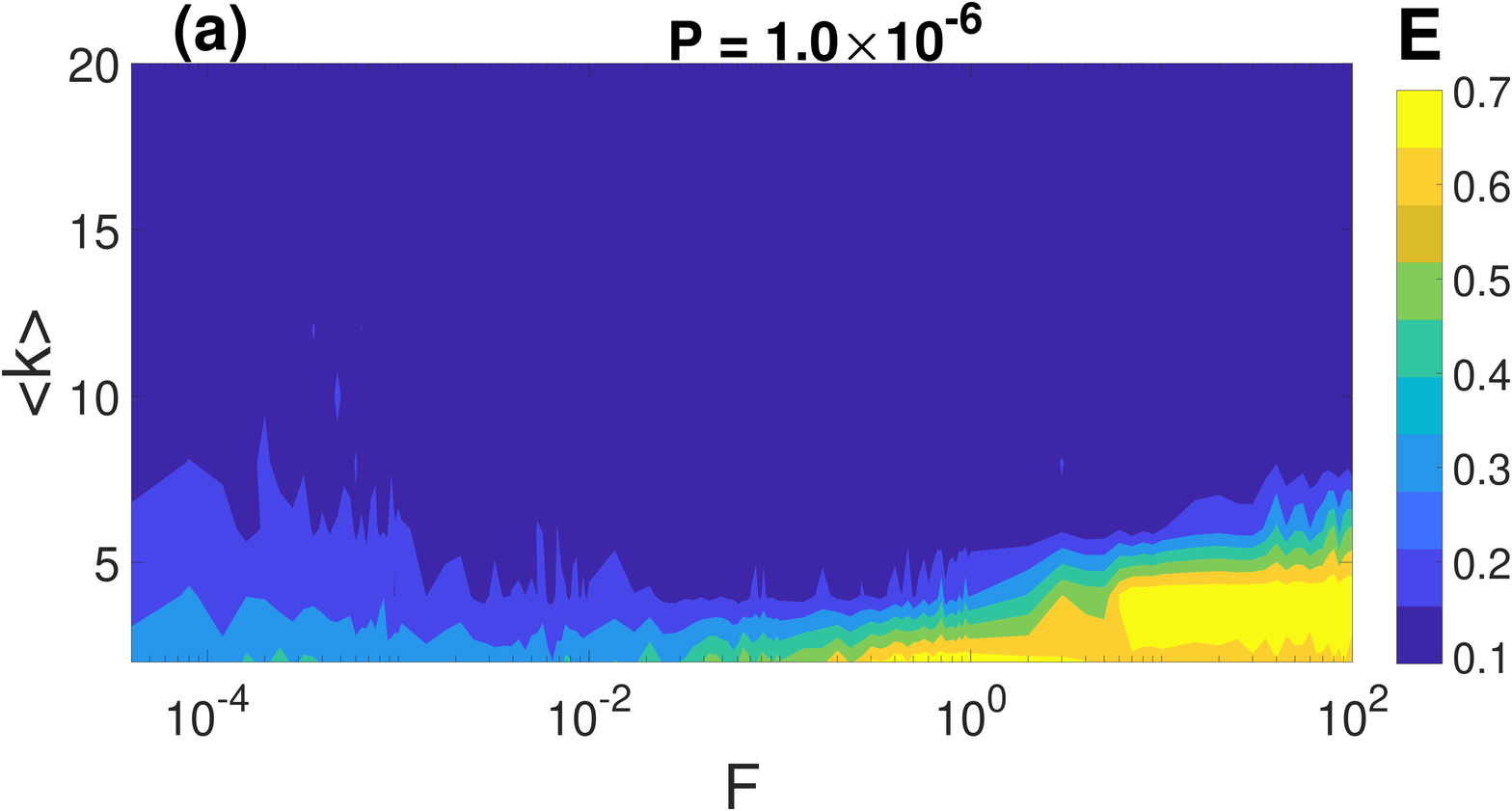}\includegraphics[width=6.0cm,height=4.0cm]{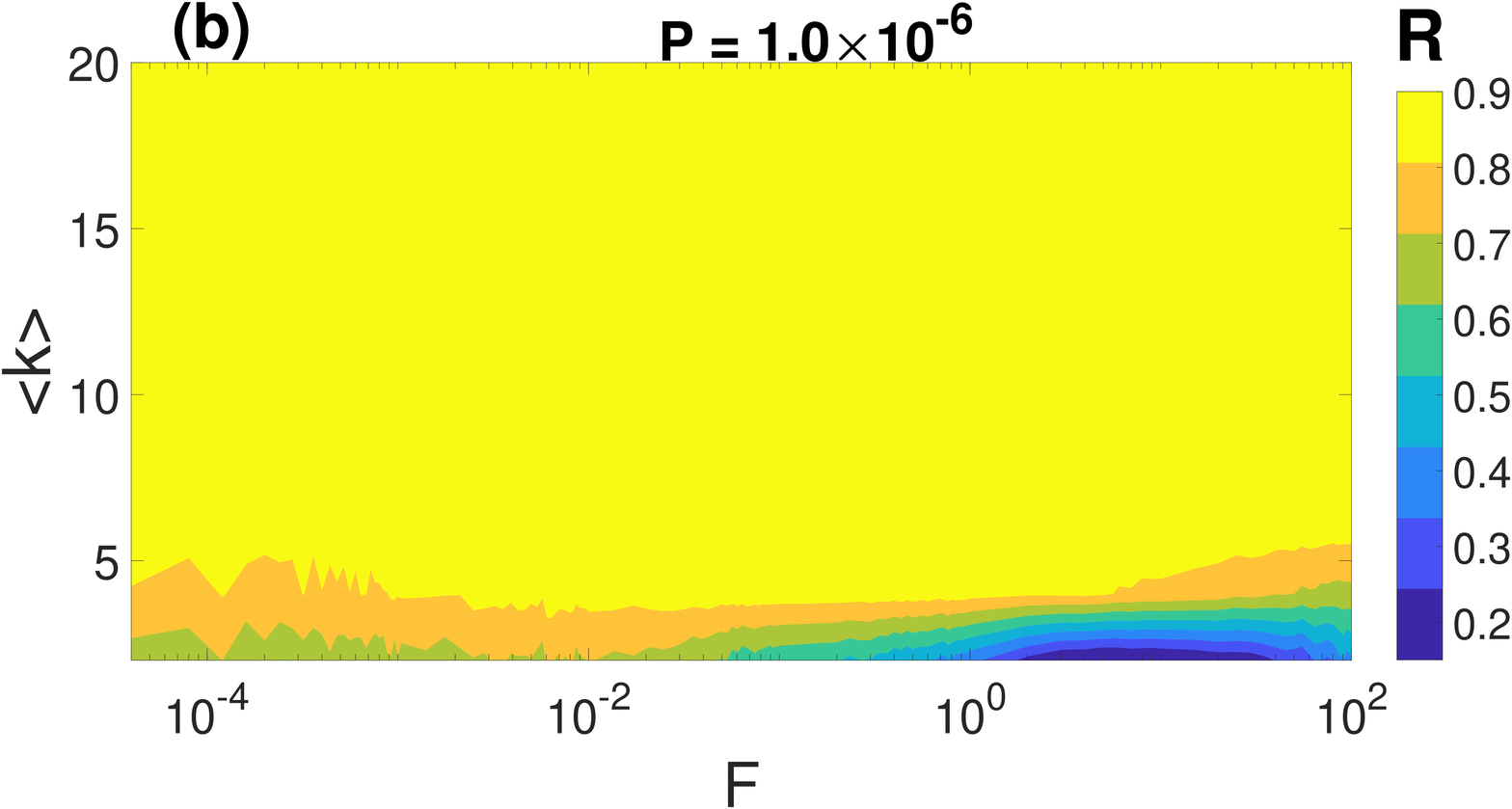}\\
\includegraphics[width=6.0cm,height=4.0cm]{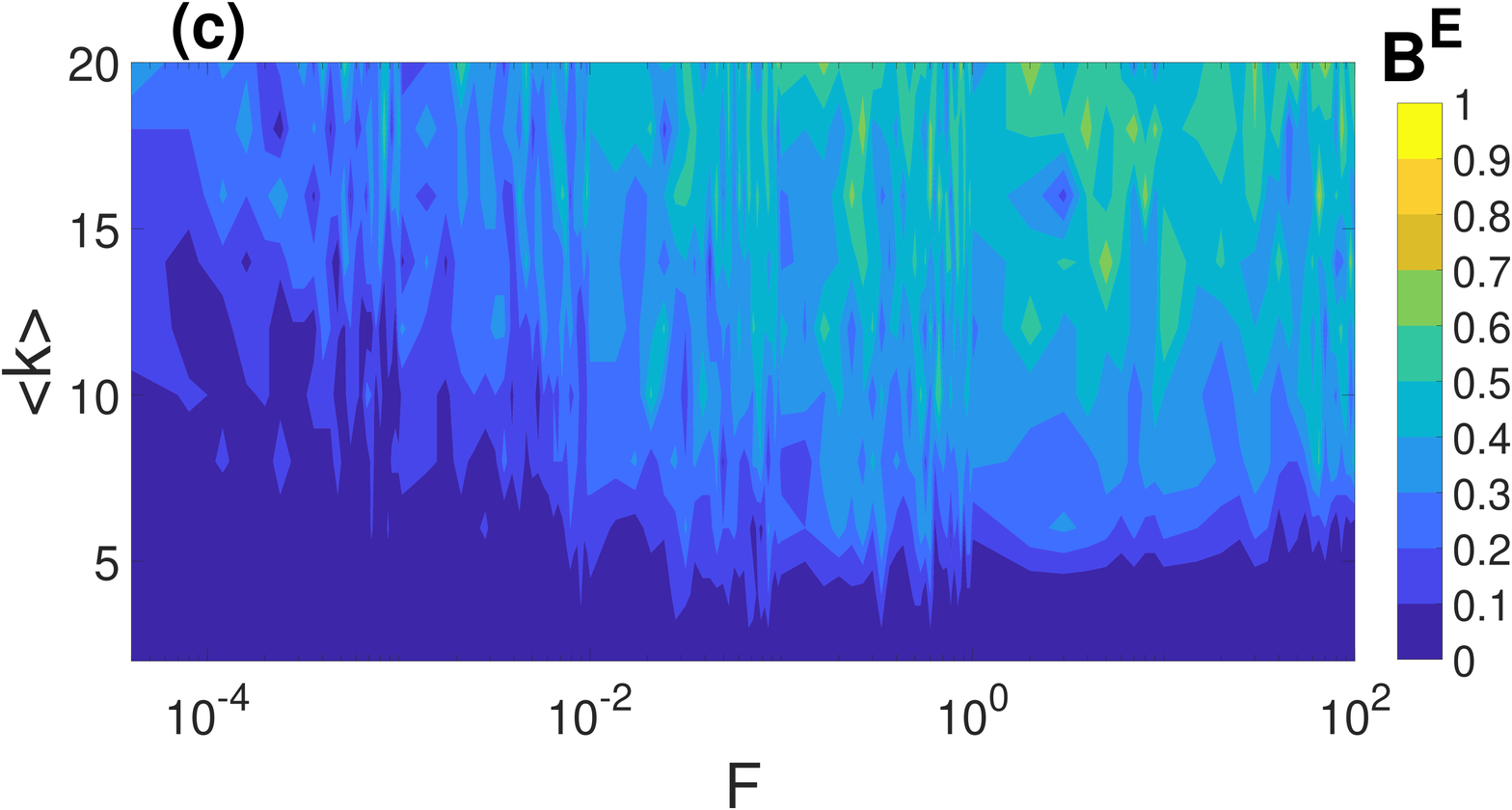}\includegraphics[width=6.0cm,height=4.0cm]{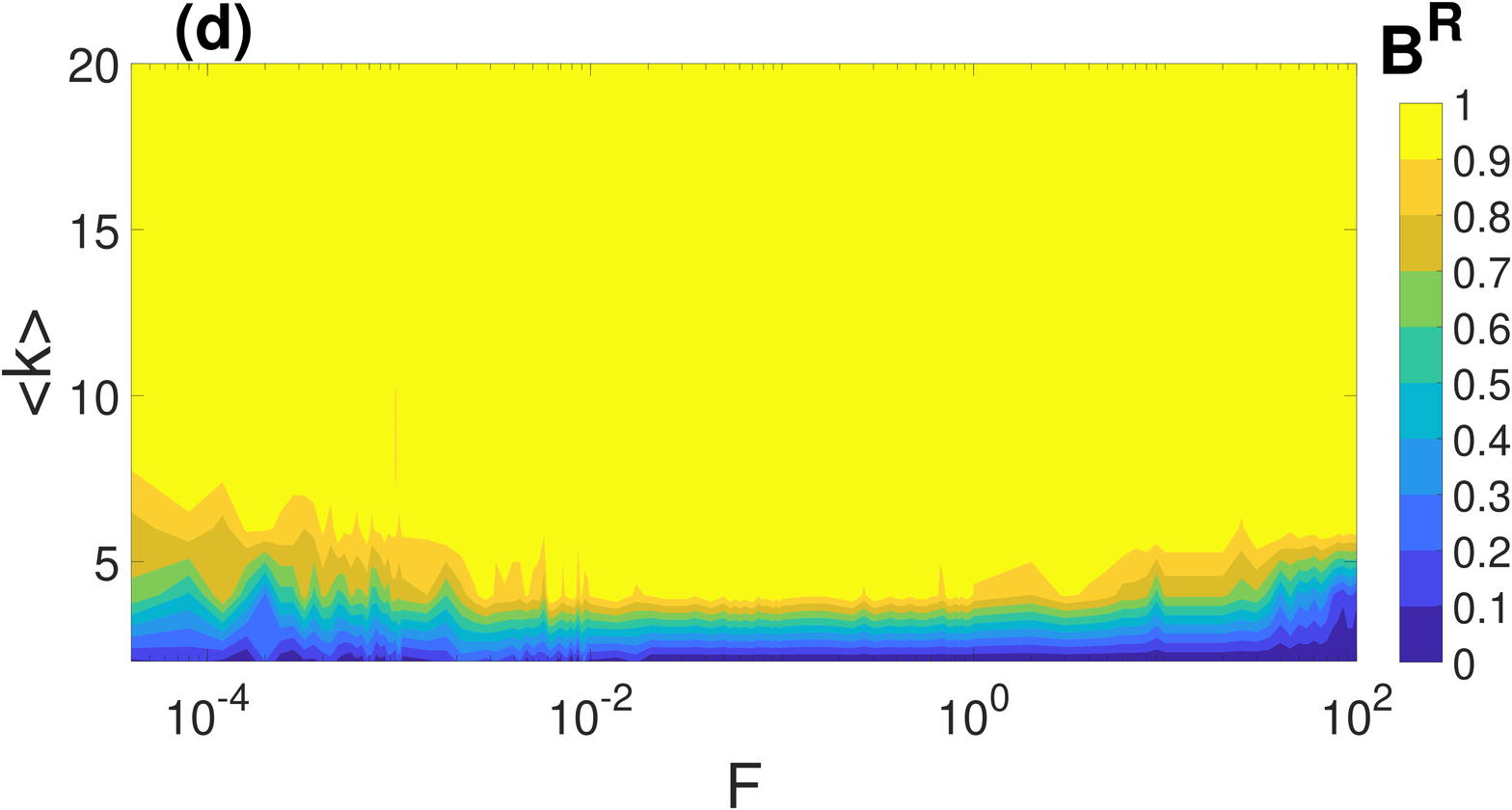}
\caption{Variation in the degree of synchronization and the corresponding global stability w.r.t. $\langle k \rangle$ and $F$ in a small-world network with an optimal STDP parameter value $P$. \textbf{(a)} and \textbf{(c)}: degree of CS and the corresponding basin stability measure. \textbf{(b)} and \textbf{(d)}: degree of PS and the corresponding basin stability measure. Parameter values: $\beta= 0.25$, $\tau_c=3.0$, $N=100$.}
\label{fig:4}
\end{figure*}

In Figs. \ref{fig:4}\textbf{(a)} and \textbf{(b)}, we depict the variations in the degree of CS and PS, respectively, as a function of  $\langle k \rangle$ and $F$ in a small-world network ($\beta=0.25$) at the optimal value of $P$ indicated. The results suggest that higher values of the average degree connectivity  ($\langle k \rangle>8$ for CS and $\langle k \rangle>5$ for PS) yield a high degree of CS and PS, irrespective of the rewiring frequency $F$. This behavior can be explained by the fact that with higher values of $\langle k \rangle$, the network becomes denser, leading to more interactions between the connected neurons which facilitate their global synchronization. 
As the small-world network becomes sparser ($\langle k \rangle<8$ for CS and $\langle k \rangle<5$ for PS), the degree of both forms of synchronization decreases, especially when the synapses switch more rapidly ($F\ge10^{-1}$).

In Figs. \ref{fig:4}\textbf{(c)} and \textbf{(d)}, we present the basin stability measures of CS and PS corresponding to Figs. \ref{fig:4}\textbf{(a)} and \textbf{(b)}, respectively. Figure \ref{fig:4}\textbf{(c)} indicates that the highest degree of CS ($E\lessapprox0.1$) obtained at higher values of $F\ge10^{-1}$ and $\langle k \rangle>8$ is more stable than in the rest of the $F-\langle k \rangle$ plane for above-prescribed phase space region. Meanwhile, Fig. \ref{fig:4}\textbf{(d)} indicates that (i) PS is fully stable for all values of $F$ and average degree connectivity $\langle k \rangle>8$ (ii) PS is more stable than CS, since max $B^R(=1)$ $>$ max $B^E(=0.8)$.  

\begin{figure*}
\centering
\includegraphics[width=6.0cm,height=4.0cm]{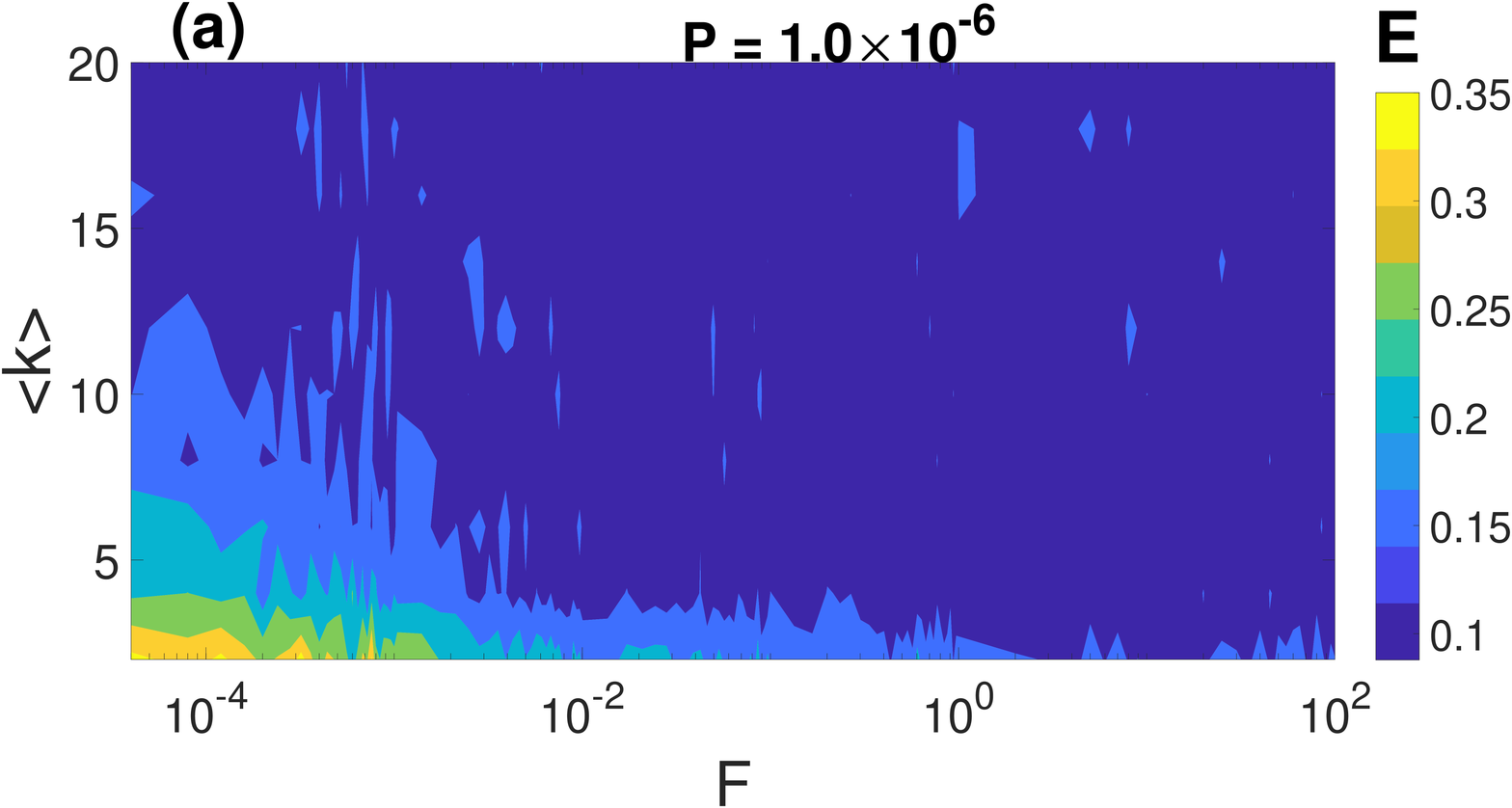}\includegraphics[width=6.0cm,height=4.0cm]{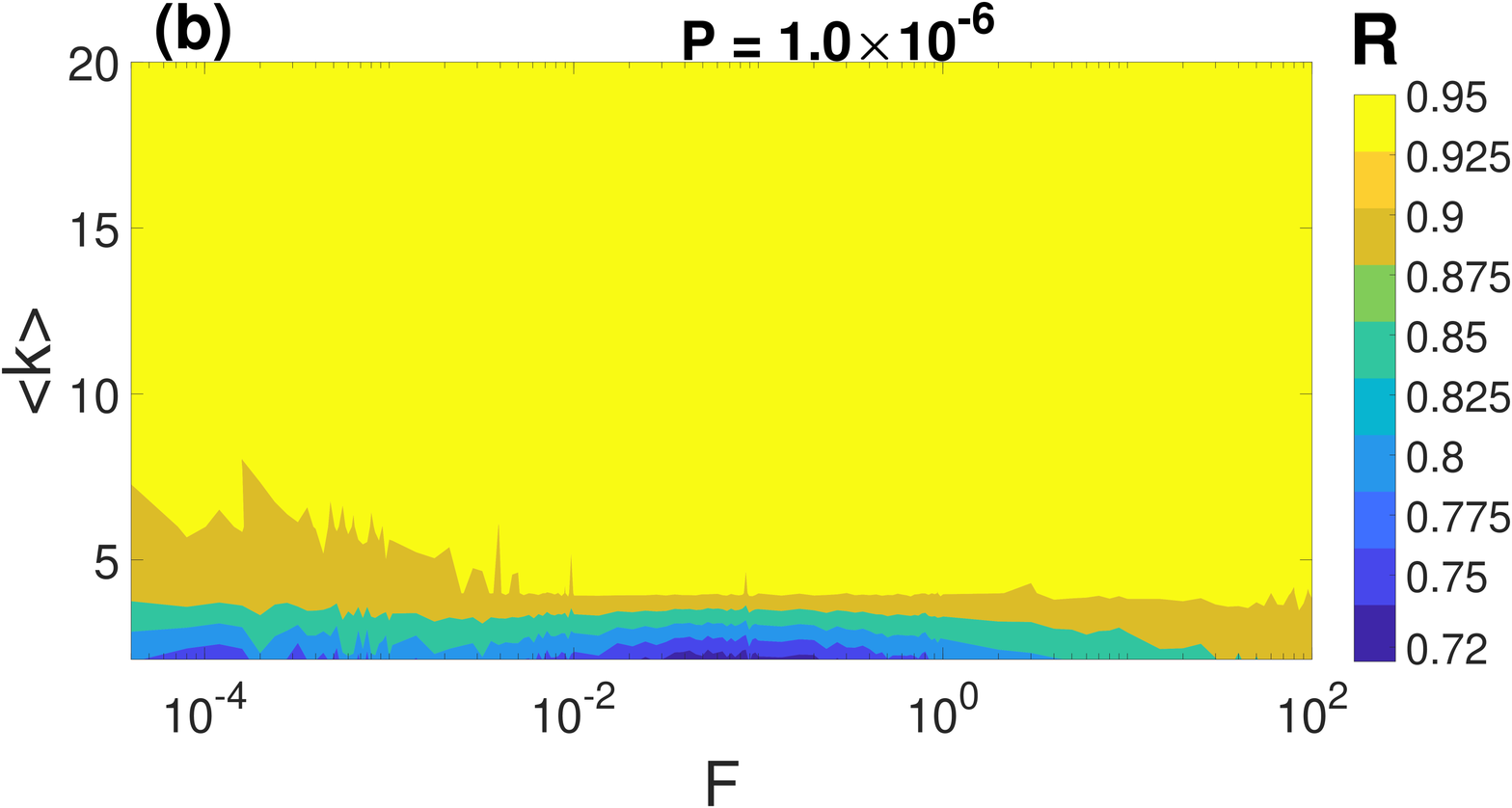}\\
\includegraphics[width=6.0cm,height=4.0cm]{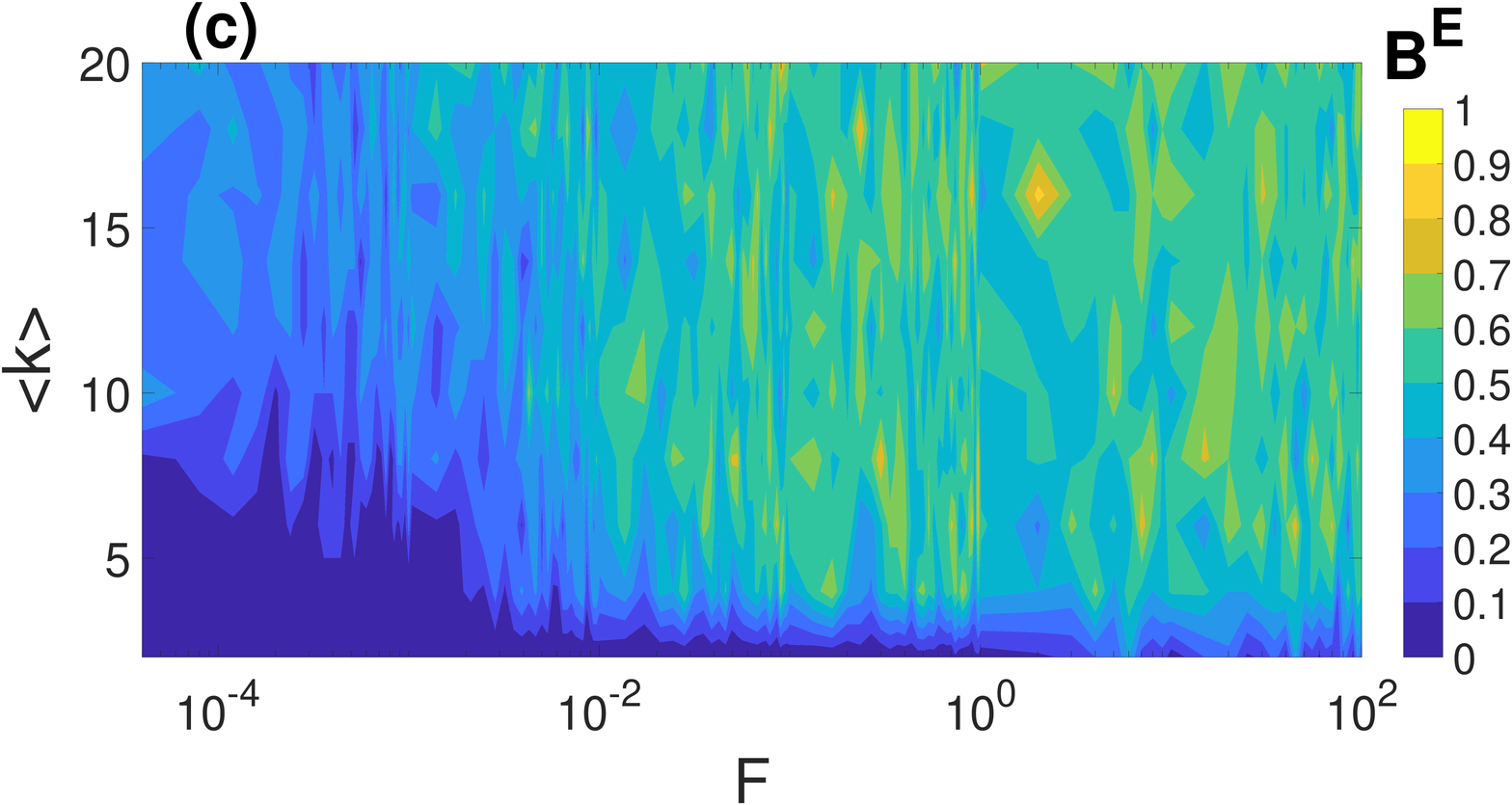}\includegraphics[width=6.0cm,height=4.0cm]{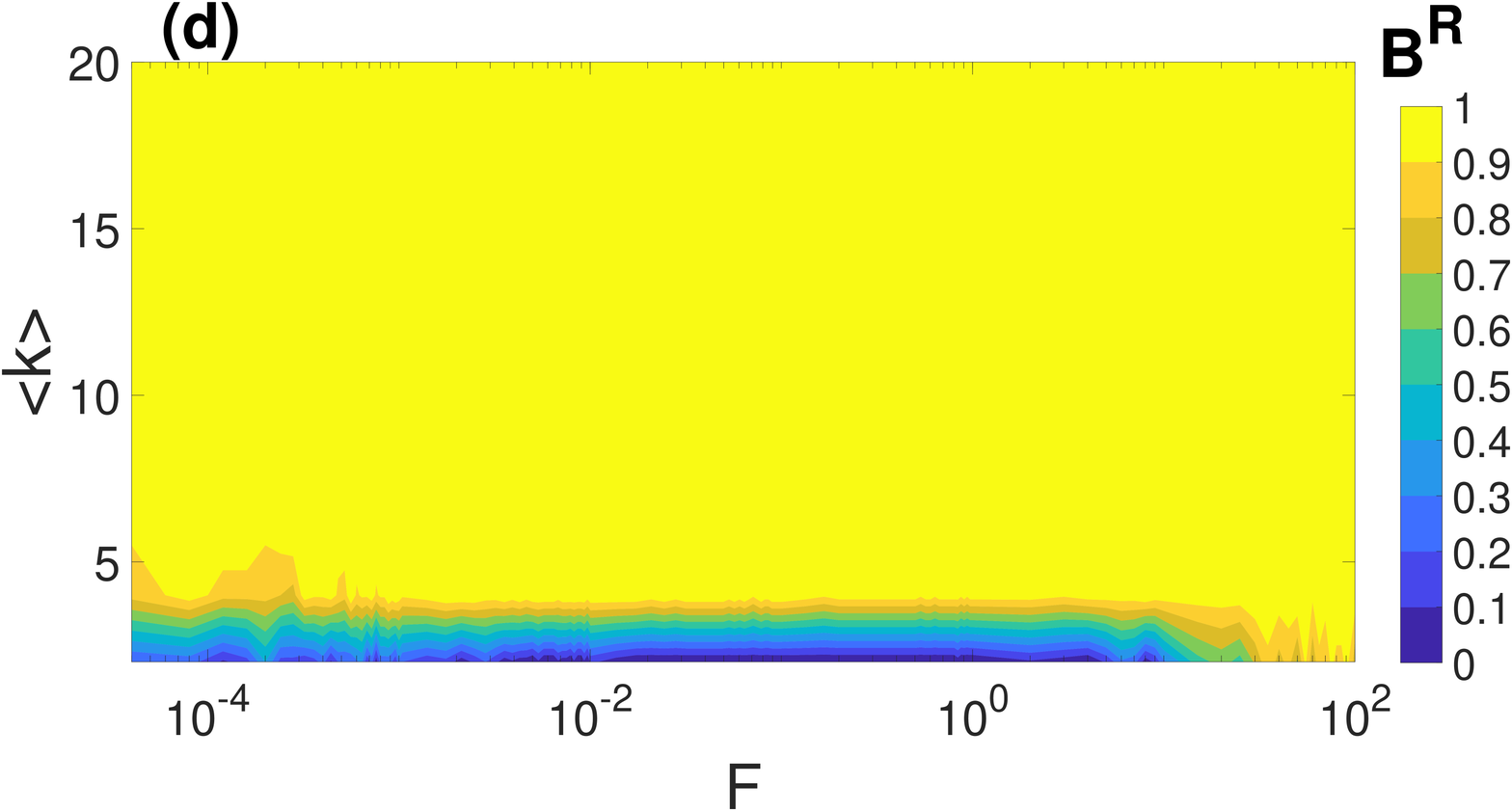}
\caption{Variation in the degree of synchronization and the corresponding global stability w.r.t. $\langle k \rangle$ and $F$ in the completely random network with an optimal STDP parameter value $P$. \textbf{(a)} and \textbf{(c)}: degree of CS and the corresponding basin stability measure. \textbf{(b)} and \textbf{(d)}: degree of PS and the corresponding basin stability measure. Parameter values: $\beta= 1.0$, $\tau_c=3.0$, $N=100$.}
\label{fig:5}
\end{figure*}

In the case of the random network ($\beta=1.0$) shown in Figs. \ref{fig:5}, firstly, we observe in Fig. \ref{fig:5}\textbf{(a)} that higher values of $F$ $(\ge10^{-2})$ increases the degree of CS irrespective of the value of $\langle k \rangle$, while lower values of $F$ $(<10^{-2})$ deteriorate the degree CS for lower values of $\langle k$ $\rangle(<8)$. This is in contrast with a small-world network in Fig. \ref{fig:4}\textbf{(a)}, where lower values of $F$ $(<10^{-2})$ enhance the degree of CS, especially at higher values of $\langle k$ $\rangle(>8)$. 
In Fig. \ref{fig:5}\textbf{(b)}, we observe that for lower  $\langle k$ $\rangle(<8)$ lower values of $F(<1)$, the degree of PS deteriorates. But unlike with degree of PS in Fig. \ref{fig:4}\textbf{(b)}, which decreases for $\langle k$ $\rangle(<5)$ and $F$ $(>1)$, the degree of PS in Fig. \ref{fig:5}\textbf{(b)} slightly increases for these same ranges of parameter values. 

Secondly, it can be seen that the degrees of CS and PS are significantly higher in the random network in Fig. \ref{fig:5} than in the small-world network in Fig. \ref{fig:4}.  In Figs. \ref{fig:5}\textbf{(c)} and \textbf{(d)}, we present the basin stability measures of CS and PS corresponding to Figs. \ref{fig:5}\textbf{(a)} and \textbf{(b)}, respectively. It is evident that CS and PS are more stable in the random than in the small-world network depicted in Figs. \ref{fig:4}\textbf{(c)} and \textbf{(d)}. These behaviors can be explained by the fact that in a random network, neurons interact, on average, with as many nearest and as distant neighbors, while in the small-world network (with $\beta=0.25$), most of the neurons interact only with their nearest neighbors and a relatively few distant neighbors. These fewer interactions in the small-world network reduce the degree of synchronization.

\marius{In summary, in the $\langle k \rangle-F$ parameter plane, lower values of $F$ and higher values of  $\langle k \rangle$ yield higher and more stable degrees of CS and PS in small-world networks. While the higher values of $F$ and higher values of $\langle k \rangle$ yield higher and more stable degrees of CS and PS in the random network. }


\subsection{Combined effect of $F$ and $\beta$ at the optimal $P$}
\begin{figure*}
\centering
\includegraphics[width=6.0cm,height=4.0cm]{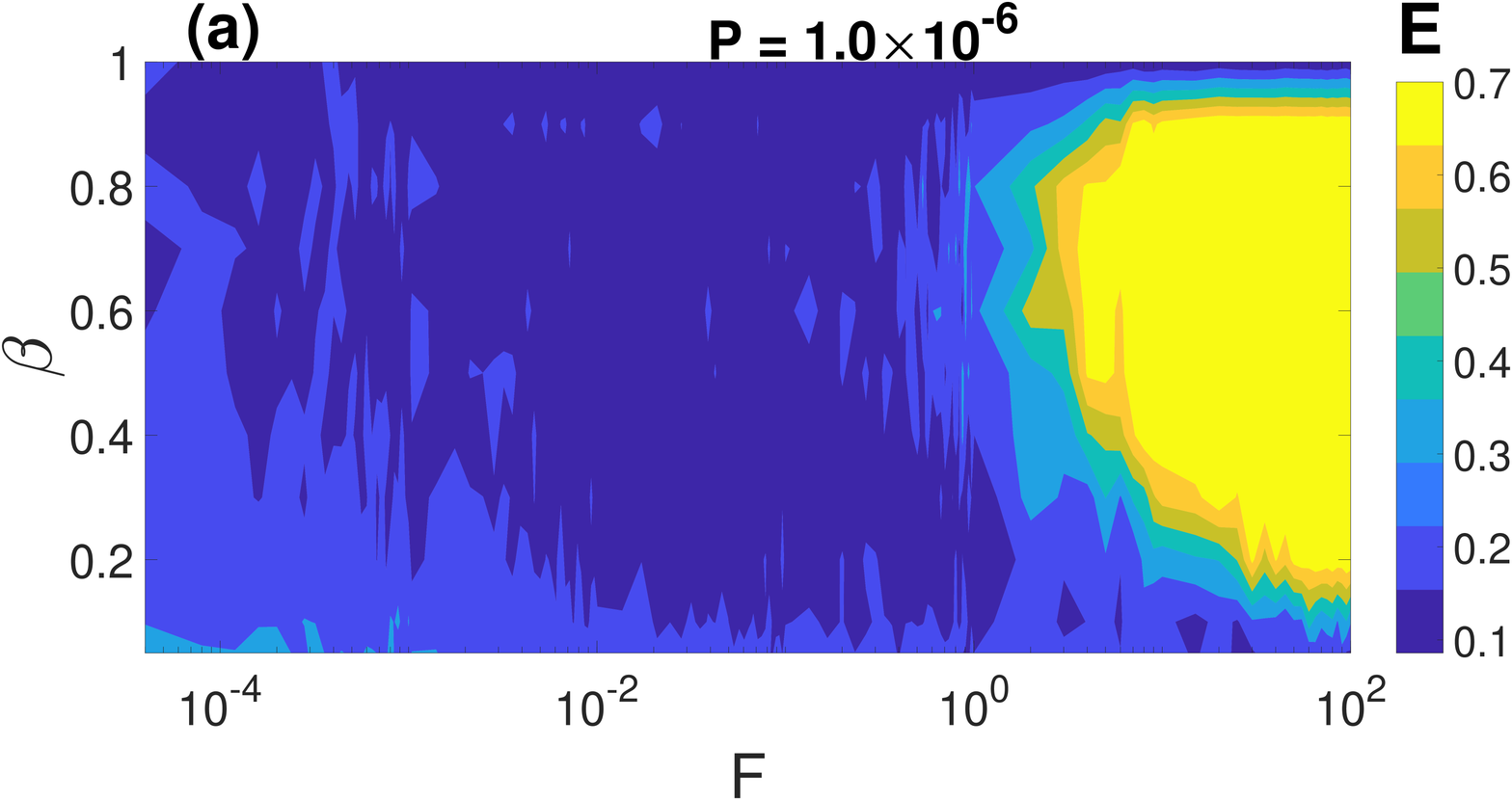}\includegraphics[width=6.0cm,height=4.0cm]{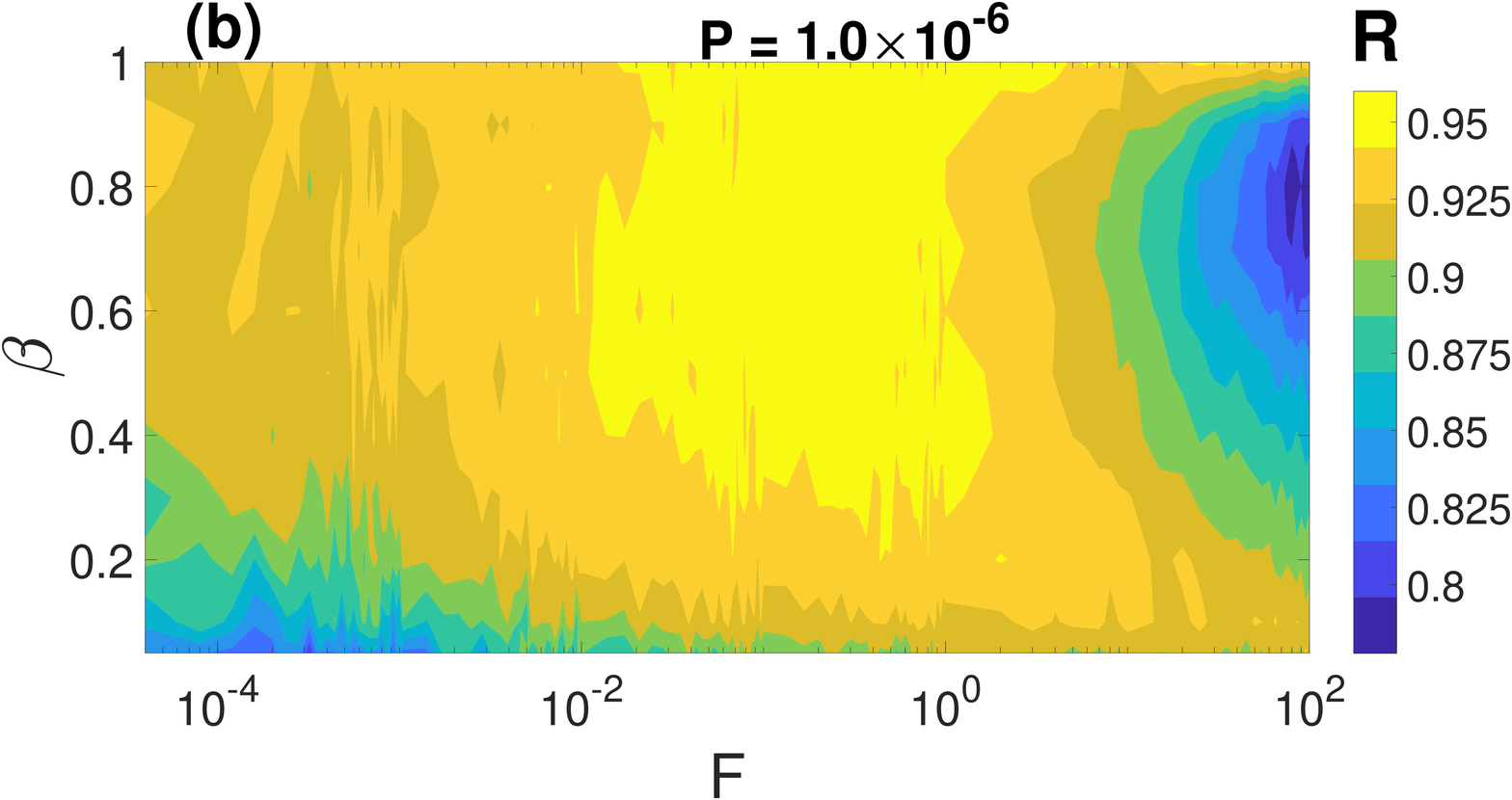}\\
\includegraphics[width=6.0cm,height=4.0cm]{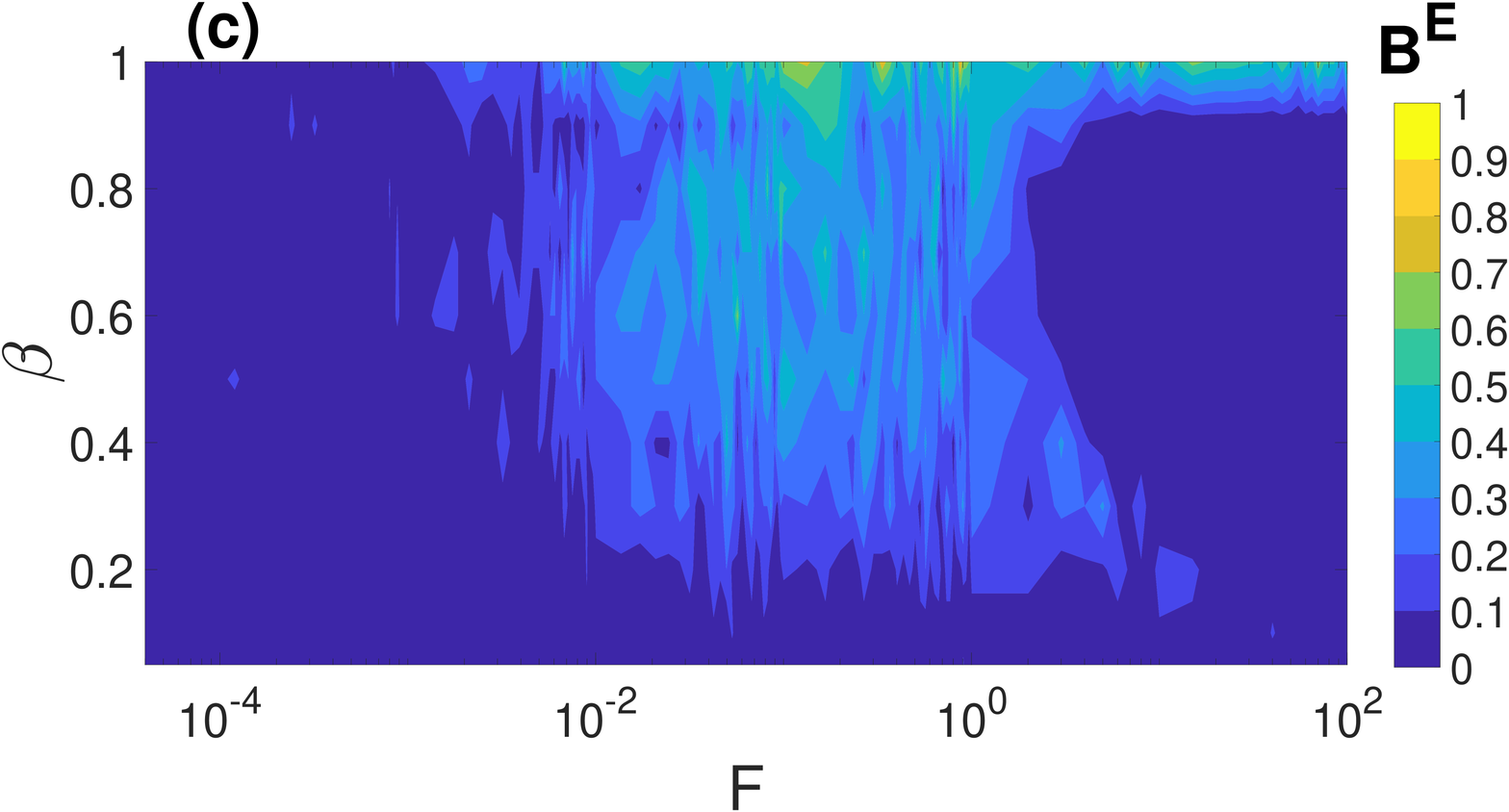}\includegraphics[width=6.0cm,height=4.0cm]{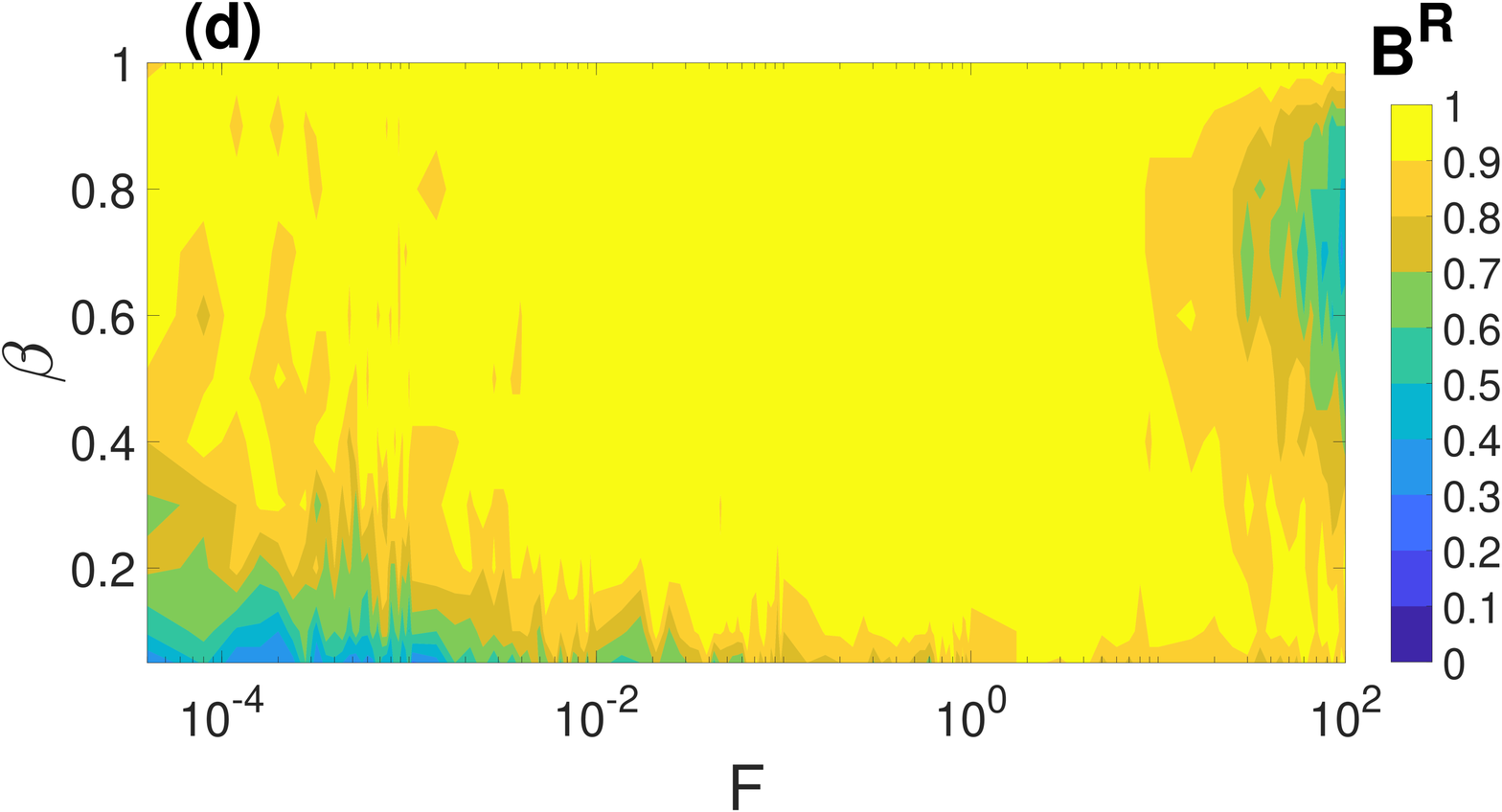}
\caption{Variation in the degree of synchronization and the corresponding global stability w.r.t. $\beta$ and $F$ with an optimal STDP parameter value $P$. \textbf{(a)} and \textbf{(c)}: degree of CS and the corresponding basin stability measure. \textbf{(b)} and \textbf{(d)}: degree of PS and the corresponding basin stability measure. Parameter values:  $\langle k \rangle=5$, $\tau_c=3.0$, $N=100$.}
\label{fig:6}
\end{figure*}

In Figs. \ref{fig:6}\textbf{(a)} and \textbf{(b)}, we show the variations in the degree of CS and PS, respectively, as a function of $\beta\in[0.05,1]$ and $F$ at the optimal value of $P$ indicated. It can be seen that the degrees of CS and PS are relatively low for (i) small-world networks built with a low rewiring probability (i.e., $\beta<0.1$) and have slowly switching synapses (i.e., $F<10^{-3}$) and (ii) for almost all small-world networks with rapidly switching synapses (i.e., $F>1$). For the random network (i.e., when $\beta=1$), the degrees of CS and PS stay relatively high irrespective of $F$.

In Figs. \ref{fig:6}\textbf{(c)} and \textbf{(d)}, we present the basin stability measures of CS and PS corresponding to Figs. \ref{fig:6}\textbf{(a)} and \textbf{(b)}, respectively.
It can be observed that the CS is more stable for more small-world networks with a higher number of random shortcuts (i.e., higher rewiring probability $\beta>0.4$) and intermediate rewiring frequencies (i.e., $10^{-2}<F<10^{0}$). For the case of a completely random network (i.e., $\beta=1$), we have more stable CS for a wider range of the rewiring frequency (i.e., $10^{-2}<F<10^{2}$). The degree of PS in Fig. \ref{fig:6}\textbf{(b)} shows similar behavior.  Comparing Figs. \ref{fig:6}\textbf{(a)} and \textbf{(b)}, we see that PS is more stable than CS both in terms of the size of the region where $B^E$ and $B^R$ achieve their maximum values and of the actual maximum values of $B^E$ and $B^R$.   

\marius{ In summary, in the $\beta-F$ parameter plane, higher values of $\beta\in[0.05,1]$ and intermediate values of $F$ yield a higher and more stable degree of CS and PS (i.e., random network yields better synchronization than small-world networks). }

\section{Conclusions}\label{Sec. V}
\marius{
This paper investigated the properties of two important phenomena, complete synchronization (CS) and phase synchronization (PS), in adaptive small-world and random neural networks. These networks were driven by two adaptive rules: Spike-Timing-Dependent Plasticity (STDP) and homeostatic structural plasticity (HSP). Our study yielded valuable insights into the factors that significantly affect the degree and stability of CS and PS. We found that various parameters, including the potentiation rate parameter $P$ for STDP, the rewiring frequency parameter $F$ for HSP, and the network topology parameters such as synaptic time delay $\tau_c$, average degree connectivity $\langle k \rangle$, and rewiring probability $\beta$, play a crucial role in shaping the dynamics and stability of CS and PS.
}

\marius{
Our results consistently demonstrated that PS exhibits greater stability compared to CS. This observation is particularly significant because precise spike timing is known to be crucial for information processing in neural systems \cite{pei1996noise}. The greater stability of PS, as indicated by the basin stability measure, may explain why neurons rely on the precise timing of spikes to encode information rather than the trace of the voltage (represented by the actual values of the voltage $v$), which is used to evaluate the degree of CS through the error $E$.
}

\marius{
Furthermore, recent experiments have shown that the modulation of STDP can be influenced by signaling molecules such as acetylcholine \cite{brzosko2019neuromodulation}. Additionally, advances in Neuroscience research have made it possible to manipulate synapse control in the brain using drugs that affect neurotransmitters \cite{pardridge2012drug} or optical fibers to stimulate genetically engineered neurons selectively \cite{packer2013targeting}. Consequently, our findings hold practical implications for optimizing neural information processing through synchronization in experimental settings and designing artificial neural circuits that enhance signal processing through synchronization.
}


\begin{appendices}\label{Appendix}
\addcontentsline{toc}{section}{Appendix}
\section*{Appendix}

\begin{table}[H]
\caption{Definition of notations used in the Algorithm}\label{tab2}
\begin{tabular}{|c|c|}
\hline
 $N$    &   network size  \\
 \hline
 $t$    &   time  \\
 \hline
 $T$    &   total integration time \\
 \hline
 $X_i(t)$      &  set  of variables $\{ v_i(t),\:w_i(t),\:\phi_i(t)\}$ in Eqs.\eqref{eq:1}\\
 \hline
 $Q$    &  total number of realizations\\
 \hline
 $F$   &   rewiring frequency of synapses   \\
 \hline
 $F_{max}$  & max rewiring frequency  \\
 \hline
$P$  &  STDP control parameter  \\
 \hline
 $P_{min}$ &  min of $P$ \\
 \hline
$P_{max}$   &  max of $P$   \\
 \hline
 $\ell_{ij}(t)$  &  adjacency  matrix of synapses \\
  \hline
 $g_{ij}(t)$  &    synaptic weights     \\
     \hline
  $\beta$  &  rewiring probability in Watts-Strogatz algorithm  \\
  \hline
 $t^n_i$    &   $n^{th}$ spike time of the $i^{th}$ neuron      \\
     \hline
 $r_q$ &   order parameter of the $q^{th}$ realization        \\
     \hline
 $e_q$   &    synchronization error of the $q^{th}$ realization      \\
  \hline
  $g_q$   &   mean  synaptic weight of the $q^{th}$ realization    \\
      \hline
  $B^{E}_{q}$  &   basin stability of CS for the $q^{th}$ realization    \\
      \hline
 $B^{R}_{q}$  & basin stability of PS for the $q^{th}$ realization  \\
      \hline
      $Tol_{E}$     &  tolerance value of $e_q$ for CS   \\
       \hline
   $Tol_{R}$     &  tolerance value of $r_q$ for PS     \\
          \hline
   $C_{E}$     &  \# of initial conditions that finally arrive at $Tol_{E}$        \\
          \hline
   $C_{R}$     &   \# of initial conditions that finally arrive at $Tol_{R}$       \\
       \hline
  $E$     &  average synchronization error over $Q$         \\
        \hline
  $R$   &  average order parameter over $Q$        \\
        \hline
  $G$  &  average of mean synaptic weights over $Q$         \\
          \hline
   $B^{E}$ &  basin stability measure for CS   \\
          \hline
  $B^{R}$  & basin stability measure for PS    \\
 \hline
\end{tabular}
\end{table}

\begin{algorithm}
\KwInput{$N$, $T$, $Q$, $F$, $P$, $\beta$}
\KwOutput{$E, G,  R, B^{E}, B^{R}$}
$P \gets P_{min}$ \tcp*{Initialize the adjusting rate parameter}
\While{$ P \leq P_{max}$ }  { 
$F\leftarrow 0$ \tcp*{Initialize the rewiring frequency}
	\While{$ F \leq F_{max}$ } {
		\For{$q \in  1,2,\dots,Q$}{
			Init $X_i(t)\:,\:\ell_{ij}(t)$ \tcp*{Random initial conditions of ODEs and initial SW or RND network adjacency matrix}
			\For{$t \in 0,\dots,T $}{
				Integrate network of ODEs in Eq. \eqref{eq:1}\tcp*{Using the 4th order Runge Kutta method}
				Record the voltage spike times $t^n_i$\tcp*{Times $t$ at which $v_i(t)\ge v_{\mathrm{th}}=0.5$}
				\If{$\Delta t_{ij}:=t_i - t_j > 0$}  {
					$\Delta M\gets P\exp{(-\lvert\Delta t_{ij}\rvert/\tau_{p})}$ 
			\tcp*{$t_i$\:,\:$t_j$: nearest-spike times of post ($i$) \& pre ($j$) neuron} 
			}
		
				\If{$\Delta t_{ij} <0$} {
				$\Delta M \gets  -1.05 P\exp{(-\lvert\Delta t_{ij}\rvert/\tau_{d})}$ 
			}
		
			\If{$\Delta t_{ij} = 0$} {
				$\Delta M \gets  0$ 
			}
				$g_{ij}(t) \gets g_{ij}(t) + g_{ij}(t)\Delta M$\tcp*{update synaptic weights}
				$\ell_{ij}(t) \gets \widetilde{\ell_{ij}}(t) $ \tcp*{Update the adjacency matrix with  $\widetilde{\ell_{ij}}(t)$ obtained by randomly rewiring $\ell_{ij}(t)$ with frequency $F$ according to the small-world ($\beta\in(0,1)$) or random ($\beta=1$) network rewiring model}
		}
			$g_q \gets \displaystyle{\bigg \langle \frac{1}{N^2}\sum\limits_{i=1}^{N}\sum\limits_{j=1}^{N}g_{ij}(t)\bigg \rangle_t}$\;
		$e_q \gets \displaystyle{\Bigg \langle\frac{1}{N-1}\sum\limits_{i=2}^N\sqrt{(v_i-v_1)^2+(w_i-w_1)^2+(\phi_i-\phi_1)^2}\Bigg \rangle_{{t}}}$\;
		$r_q \gets \displaystyle{\bigg \langle\bigg\lvert\frac{1}{N}\sum\limits_{i=1}^N\exp{\Big[z\Big(2\pi \ell + 2\pi(t-t_{_{i}}^{n})/(t_{_{i}}^{n+1}-t_{_{i}}^{n})\Big)\Big]}\bigg\rvert\bigg \rangle_t}$\tcp*{$z=\sqrt{-1}$}
		
		$C_{E} \gets 0$\;
			\If{$e_q < Tol_{E}$} {
			$C_{E} \gets C_{E} + 1$\tcp*{compute the number of initial conditions that finally reach or exceed the tolerance for CS} 
			}
		$C_{R} \gets 0$\;
			\If{$r_q > Tol_{R}$} {
			$C_{R} \gets C_{R} + 1$\tcp*{compute  the number of initial conditions that finally reach or exceed the tolerance for PS} 
			}
			Add $e_q$ to $e$\;
			Add $g_q$ to $g$\;
			Add $r_q$ to $r$\;
		}
		$E \gets e/Q$\tcp*{Compute the  average of $E$ over $Q$ }
		$G \gets g/Q$\tcp*{Compute the average of $G$ over $Q$ }
		$R \gets r/Q$\tcp*{Compute the average of  $R$ over $Q$}
		$B^{E} \gets C_{E}/Q$\tcp*{Compute basin stability measure for CS}
		$B^{R} \gets C_{R}/Q$\tcp*{Compute basin stability measure for PS}
		$F \gets F + \Delta F$\tcp*{Increment the HSP control parameter}
	} 
$P \gets P + \Delta P$  \tcp*{Increment the STDP control parameter}
}
\caption{Flow of control in the simulations}
\end{algorithm}

\end{appendices}
 \newpage

\section*{Acknowledgments}
 MEY acknowledges the support from the Deutsche Forschungsgemeinschaft (DFG, German Research Foundation) -- Project No. 456989199 and the Neuromod Institute of the Universit{\'e} C{\^o}te d'Azur, Sophia Antipolis, France, and the warm hospitality of the MathNeuro Inria Project-Team. SR acknowledges support from Elkartek 2023 via grant ONBODY no. KK-2023/00070.

\section*{Author Contributions} MEY conceptualized the study and did the numerical simulations. All authors contributed equally to the result analysis and
manuscript writing. All authors read and approved the final manuscript.

\section*{Data availability statement} 
The simulation data supporting this study's findings are available within the article.

\section*{Declaration of competing interest}
The authors declare that there is no conflict of interest in relation to this article.


\end{document}